\documentclass[pop,superscriptaddress,showpacs,amsmath,amssymb,reprint]{revtex4-1}
\usepackage{graphicx,dcolumn,bm,amsmath}
\usepackage[normalem]{ulem}

\newcommand{\beq}{\begin{equation}}
\newcommand{\enq}{\end{equation}}

\newcommand\Real{\mbox{Re}} 

\newcommand{\e}{\eqref}

\newcommand \bfr{{\bf r}}
\newcommand \bfk{{\bf k}}

\newcommand{\p}{\partial}

\begin{document}

\title{
Langmuir wave filamentation in the kinetic regime. II. Weak and Strong Pumping of Nonlinear Electron Plasma Waves as the Route to Filamentation
}

\author{Denis A. Silantyev}
\affiliation{Department on Mathematics and Statistics, University of New Mexico, New Mexico 87131, USA}
\author{Pavel M. Lushnikov}
\email{plushnik@math.unm.edu}
\affiliation{Department on Mathematics and Statistics, University of New Mexico, New Mexico 87131, USA}
\author{Harvey A. Rose}
\affiliation{Theoretical Division, Los Alamos National Laboratory,
  MS-B213, Los Alamos, New Mexico, 87545}
\affiliation{New Mexico Consortium, Los Alamos, New Mexico 87544, USA}

\date{
\today
}

\begin{abstract}
We consider two kinds of pumped Langmuir waves (LWs) in the kinetic regime, $k\lambda_D\gtrsim0.2,$  where $k$ is the LW wavenumber and $\lambda_D$ is the Debye length, driven to finite amplitude by a coherent external potential whose amplitude is either weak or strong. These dynamically prepared nonlinear LWs develop a transverse (filamentation) instability whose nonlinear evolution destroys the LW's transverse coherence. Instability growth rates in the weakly pumped regime are the same as those of BGK modes considered in Part I \cite{SilantyevLushnikovRosePOP2017}, while strongly pumped LWs  have higher filamentation grow rates.
\end{abstract}
\maketitle

\section{Introduction}

Propagation of intense laser beam in high temperature plasma relevant for the inertial confinement fusion results in significant loss of laser energy to stimulated Raman back-scatter (SRS)  \cite{GoldmanBoisPhisFL1965} producing the electromagnetic waves at different frequency and  Langmuir waves \cite{KrallTrivelpiece1973} (LW).  If
 the LW wavenumber $k$ satisfies, $k\lambda_D\gtrsim0.2$,  the ``kinetic" regime, then  kinetic effects related to electron trapping \cite{ManheimerFlynnPhysFL1971,DewarPhysFL1972,MoralesNeilPRL1972} become important \cite{RosePOP2005,KlineMontgomeryBezzeridesCobblePRL2005,KlineMontgomeryYin2006,WinjumFahlenMoriPOP2007,ChangDodinPOP2015}, where  $\lambda_D$ is the Debye length.  LW filamentation  in the kinetic regime  saturates SRS \cite{YinAlbrightBowersDaughtonRosePOP2008,YinAlbrightRoseBowersPOP2009} by reducing the LW's coherence.

In Part I \cite{SilantyevLushnikovRosePOP2017} of this series we addressed LW filamentation in  the kinetic regime by studying both analytically and  through 2+2D (two spatial dimensions and two velocity dimensions) spectral Vlasov simulations, the transverse instability of the special class of one-dimensional (1D) Bernstein-Greene-Kruskal (BGK) modes \cite{BernsteinGreeneKruskal1957}. That class approximates  the adiabatically slow creation of BGK modes by SRS. In this paper  we take an alternative approach by dynamically preparing BGK-like initial conditions  through either weak or strong SRS-like pumping. We found that these 1D BGK-like solutions obtained via weak pumping have the same  transverse instability growth rate as BGK modes of Part I suggesting a universal mechanism for  kinetic saturation of SRS in laser-plasma interaction experiments. We found that strong pumping (compared to weak pumping) results in further increase of the growth rate of the transverse instability thus speeding up LW filamentation. We also compare the result of our numerical simulations to the corresponding results in Ref. \cite{BergerBrunnerBanksCohenWinjumPOP2015,BanksPrivate2016}.

The paper is organized as follows. Section \ref{sec:Basicequations} introduces Vlasov-Poisson equation with external pumping imitating SRS.
Section \ref{sec:1DEPW} describes our method of producing BGK-like modes by both weak pumping (Section \ref{sec:AdiabaticEPW}) and strong pumping  (Section \ref{sec:FastEPW}).
Section \ref{sec:TransverseInstability} provides the analytical expressions on the growth rate of transverse instability of  BGK-like modes.
In Section \ref{sec:2DPumpedEPW} is devoted to results of numerical 2+2D Vlasov simulations and comparison with the theory.
Section \ref{sec:Simulationsettings} outlines to the settings of our Vlasov simulations and numerical spectral methods used.
Section \ref{sec:transverseBGKlike} addresses  transverse instability of BGK-like modes created by pumping.
Section \ref{sec:2DComparison} provides a comparison of transverse instability of BGK-like modes with BGK modes of Part I.
In Section \ref{sec:Conclusion} the main results of the paper are discussed.

\section{Basic equations}
\label{sec:Basicequations}

The Vlasov equation for the phase
space distribution function $f(\bfr,{\bf v},t)$, in units such that electron mass $m_e$ and charge $e$ are
normalized to unity, the spatial coordinate $\bfr=(x,y,z)$ to the electron Debye length $\lambda_D$, the time $t$ to  reciprocal electron plasma
frequency, $1/\omega_{pe}$, \cite{RoseDaughtonPOP2011} and  the velocity ${\bf v}=(v_x,v_y,v_z)$ is normalized to the electron
thermal speed $v_e$, is
\begin{equation}
    \left\lbrace\frac{\partial }{\partial t} + {\bf v}\cdot\nabla+ {\bf E}\cdot \frac{\partial }{\partial {\bf v}} \right\rbrace f=0,
    \label{eq:vlasov}
\end{equation}
where $\bf E$ is the electric field scaled to $k_BT_e/(\lambda _De).$   Here  $T_e$ is the background electron temperature and $k_B$  is the Boltzmann constant. Magnetic field effects are ignored for clarity. Then in   the electrostatic regime
\begin{align} \label{phidef}
{\bf E}=-\nabla \Phi,
\end{align}
 with  the electrostatic potential $\Phi.$

We consider the beating of laser and SRS light as a source of LWs, idealized as  a travelling
wave sinusoidal external potential $\Phi_{ext}$, with phase speed $v_\varphi$ and
wavenumber $k_z$:
\begin{equation}
    \Phi_{ext} =\Phi_{pump} (t)\cos[k_z (z-v_\varphi t)], \ k_z = |{\bf k}|,
    \label{eq:EXTPotential}
\end{equation}
where  $\Phi_{pump} (t)$ is  prescribed.

The total electrostatic potential, $\Phi$, is given by
\begin{equation}
    \Phi =\Phi_{ext} + \Phi_{int},
    \label{eq:ElectricPotential}
\end{equation}
where the internal potential $\Phi_{int}$ is determined from Poisson's equation
\begin{align} \label{Poisson}
\nabla^2\Phi_{int}=1-\rho,
\end{align}
where the electron density $\rho$ is given by
\begin{equation}
    \rho ({\bf r},t)=\int  f({\bf r},{\bf v},t)  d{\bf v}.
    \label{eq:density}
\end{equation}
The factor $4\pi$ is absent in equation \eqref{Poisson} because of the chosen normalization and 1 in equation \eqref{Poisson} comes from the neutralizing ion background.
Equations \e{eq:vlasov}-\e{eq:density} form a closed Vlasov-Poisson system which we solve below.
%

\section{CREATION OF 1+1D BGK-LIKE SOLUTIONS BY EXTERNAL PUMPING}
\label{sec:1DEPW}
In this Section we consider the process of creation of nonlinear electron plasma waves (EPW) by external pumping. That EPW is dynamically prepared by starting from uniform in space initial conditions with Maxwellian distribution of particle velocities and applying external electric field of constant amplitude for a finite period of time to create a nonlinear EPW with the desired amplitude. We consider two types of pump. The first type is a weak pump. We found from simulations that a pump amplitude cannot be made arbitrary small (even if applied for an arbitrary large period of time) if we aim to obtain an EPW with a given finite amplitude. Then by a weak pump we mean applying as small amplitude of the pump as possible to achieve the necessary amplitude of an EPW. The second type of pump has  ten fold larger amplitude of the pump (we called it a strong pump) compared with the first type. This allows about a ten times shorter duration of pumping.  After pumping of either type is extinguished, we observe nonlinear EPWs which are not constant amplitude waves even in 1D but rather they experience small oscillations $\sim5\% $ near an average amplitude while travelling as shown in Fig. \ref{fig:1D_phi_1_t}. In that sense we  call these solutions BGK-like modes. By construction, they are the dynamically accessible nonlinear EPWs.
We perform $1+1D$ Vlasov simulations, solving Eqs. \e{eq:vlasov}-\e{eq:density} with periodic boundary conditions in phase space $(z,v_z)$, to demonstrate the properties of these EPWs.

\subsection{ Creation of BGK-like solutions by weak external pumping}
\label{sec:AdiabaticEPW}

1D BGK-like mode is prepared by starting from the spatially uniform Maxwellian distribution
\begin{equation} \label{maxw1}
f_0(v_z)=\exp(-v_z^2/2)/\sqrt{2\pi}
\end{equation}
at $t=0$ and adding the travelling external electric potential $\Phi_{ext}$ as in Eq. \e{eq:EXTPotential}
with
\begin{equation}
    \Phi_{pump}(t) =-\phi_{pump}H(T_{off}-t)
    \label{eq:PUMPPotential}
\end{equation}
where $T_{off}$ is the time when the pumping is turned off, $H(T_{off}-t)$ is the Heaviside step function ($H(T_{off}-t)=1$ for $t<T_{off}$ and $H(T_{off}-t)=0$ for $t>T_{off}$) and $v_\varphi=\omega_{LW}(k_z)/k_z$. Here $\omega_{LW}$ is the real-valued linear LW frequency (obtained using $Z$-function \cite{FriedGell-MannJacksonWyldJNF1960}, see e.g. \cite{PitaevskiiLifshitzPhysicalKineticsBook1981,NicholsonBook1983}). In this paper we work with $\omega_{LW}(k_z=0.35)=1.22095\ldots$ and $\omega_{LW}(k_z=0.425)=1.31759\ldots$.
Note that instead of  $v_\varphi=\omega_{LW}(k_z)/k_z$, we can choose e.g. $v_\varphi$ from BGK mode of Part I which is a function of $\phi_{eq}.$ We found that such a choice results in $< 10\%$ variation of the growth rate of the transverse instability for typical values of  $\phi_{eq}$ used in Section \ref{sec:transverseBGKlike} below.

Since we pump the 1st harmonic of our system in $z$-direction then we expect that the 1st Fourier harmonic amplitude $\phi_1(t)\equiv 2|\int^{L_z}_0 \Phi_{int}(z,t)\exp{(ik_zz)}dz|/L_z$ of internal electric field to be the strongest compared to other harmonics. Indeed, we observed throughout simulations that the 2nd harmonic of $\Phi_{int}$ is about 2 orders less than $\phi_{1}$, the 3rd harmonic of $\Phi_{int}$ is about one order less than 2nd and so on.

If we pump the system continuously without turning off the external pump ($T_{off}=\infty$), we observe that $\phi_1$ does not grow further than some maximum value, instead it first increases, reaches the global maximum (sometimes the global maximum is not the first local maximum), and then it decreases (in this stage $\Phi_{int}(z)$ and $\Phi_{ext}(z)$ are out of phase and the energy is being sucked out of the system by external electric field rather then being pumped into it) after which $\phi_1$ keeps oscillating with a period much longer than the bounce period $T_{bounce}=2\pi/\omega_{bounce}$, with the bounce frequency $\omega_{bounce}\approx k_z\sqrt{\phi_1}$ in dimensionless units.

\begin{figure}
\includegraphics[width=3in]{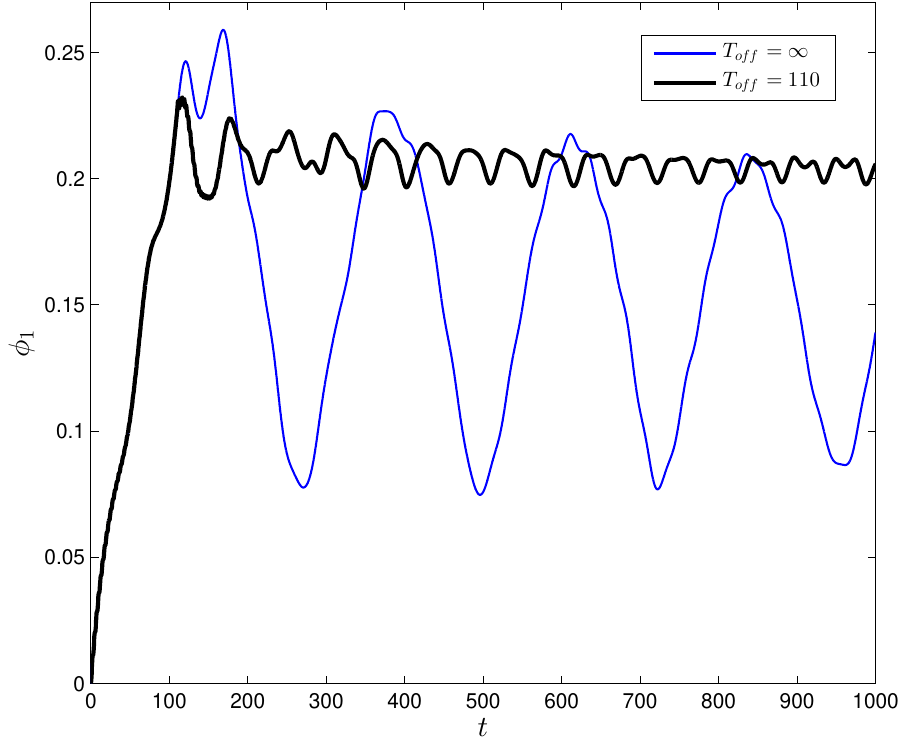}
\caption{(Color online) The evolution of $\phi_1$, the first harmonic of internal electric field, for two cases: $T_{off}=\infty$ and $T_{off}=110$. $\phi_{pump}=0.01$ and $k_z=0.35$ for the both cases.}
\label{fig:1D_phi_1_t}
\end{figure}

Fig. \ref{fig:1D_phi_1_t} shows evolution of $\phi_1(t)$ for the two cases with $T_{off}=\infty$ and $T_{off}=110$. In both cases we take $\phi_{pump}=0.01$ and $k_z=0.35.$ In the first case,   $\phi_1$ experiences the initial growth, after which it keeps oscillating with a period $T_{big}\approx230$ around an average value $\sim0.15$. Notice that the global maximin of $\phi_1(t)$ is actually the second local maximum and the duration between two local maximums (at $t\approx121$ and $t\approx169$) is $\simeq 48$ which corresponds to the bounce period $ T_{bounce}\approx2\pi/(k_z\sqrt{\phi_1})\approx 2\pi/(0.35 \sqrt{0.15}) \approx 46$.
 In the second case, when the external pump is turned off at $t=T_{off}=110$, $\phi_1$ after short transient behaviour remains almost constant ($\approx0.21$) for the rest of time experiencing  small oscillations around the average value, which we call $\phi_{eq}$.

\begin{figure}
\includegraphics[width=3in]{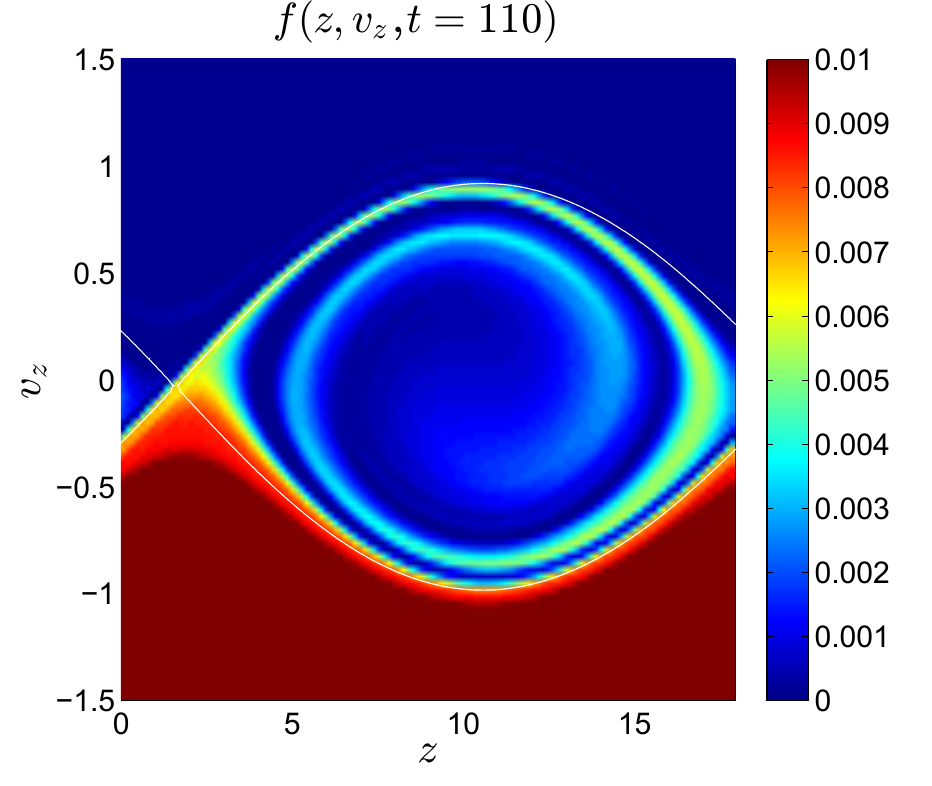}
\caption{(Color online) The density plot of $f(z,v_z,t)$ at $t=T_{off}=110$ with $\phi_{pump}=0.01$ and $k_z=0.35$. White contour marks the boundaries of the trapping region, the fraction of trapped particles is $n_{trapped}/n_{total}=0.00222$.}
\label{fig:1D_f_z_vz_t110}
\end{figure}

\begin{figure}
\includegraphics[width=3in]{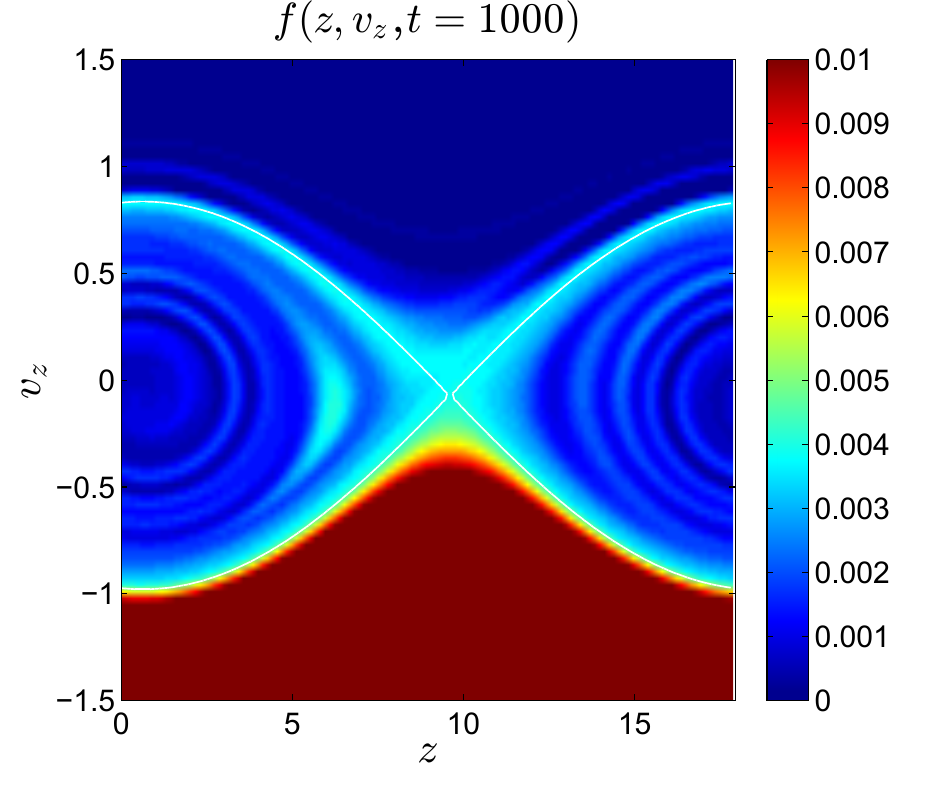}
\caption{(Color online) The density plot of the phase space distribution function $f(z,v_z,t)$ at $t=1000$. $\phi_{pump}=0.01$,$T_{off}=110$ and $k_z=0.35$. White contour marks the boundaries of the trapping region, the fraction of trapped particles is $n_{trapped}/n_{total}=0.00216$.}
\label{fig:1D_f_z_vz_t1000}
\end{figure}

\begin{figure}
\includegraphics[width=3in]{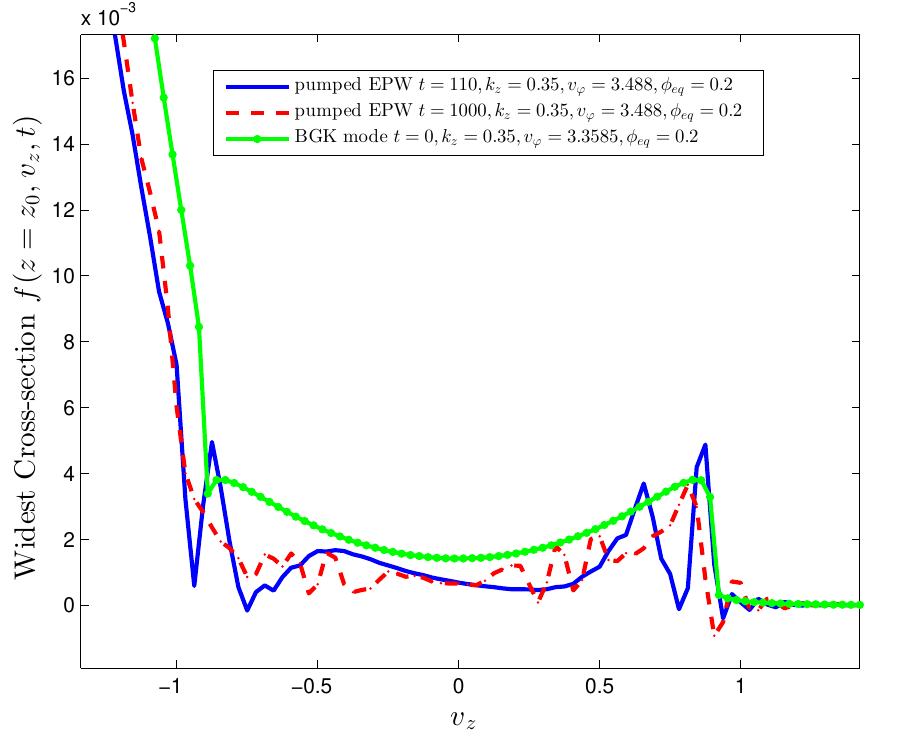}
\caption{(Color online) The widest cross-sections of $f(z=z_0,v_z,t)$ of the trapping regions at $t=110$ and $t=1000$ of the weakly pumped EPW obtained with parameters $k_z=0.35$, $v_{\varphi}=3.488$, $\phi_{pump}=0.01$, $T_{off}=110$ and resulting $\phi_{eq}=0.2$ in comparison with the widest cross-sections of the trapping region of BGK mode constructed analytically in Part I with parameters $k_z=0.35$, $\phi_{eq}=0.2$ and $v_{\varphi}=3.3585$ (according to the BGK dispersion relation). $z_0$ is chosen such that the resulting cross-sections have the maximum width.}
\label{fig:1D_Cross_sections}
\end{figure}

Figs. \ref{fig:1D_f_z_vz_t110} and \ref{fig:1D_f_z_vz_t1000} show   snapshots of the electron phase space  distribution function $f(z,v_z,t)$ around the trapping region for the simulation with $T_{off}=110$ at times $t=T_{off}=110$ and $t=1000$, respectively.   A spiral can be seen in these Figs. to develop in the trapping region with a number of revolutions $\approx t/T_{bounce}$. Fig. \ref{fig:1D_Cross_sections} shows the widest cross-sections of the  trapping region from the same times as in Figs. \ref{fig:1D_f_z_vz_t110} and \ref{fig:1D_f_z_vz_t1000}. They are also compared to the cross-section of the BGK mode of the same amplitude $\phi_{eq}=0.2$ from Part I that was constructed analytically with parameters $k_z=0.35$, $\phi_{eq}=0.2$ and $v_{\varphi}=3.3585$ (according to the BGK dispersion relation Eq. (22) in Part I). The trapping regions in Fig. \ref{fig:1D_Cross_sections} have the same width since the waves have the same amplitude while the absolute values of $f(z,v_z,t)$ are higher for the BGK mode since it has the smaller $v_{\varphi}$. These results were obtained in moving frame with the velocity $v_\varphi$. The spiral in the density distribution function of the BGK-like mode develops  increasingly smaller scale structures with time that need increasingly higher number of grid points to be resolved accurately. In our simulations these smaller scale structures are smoothed out by the presence of small hyper-viscosity (see more discussion in Section \ref{sec:2DPumpedEPW}) which is chosen to be small enough to not affect the amplitude $\phi_{eq}$ of BGK-like mode during the entire time of simulation.

\begin{figure}
\includegraphics[width=3in]{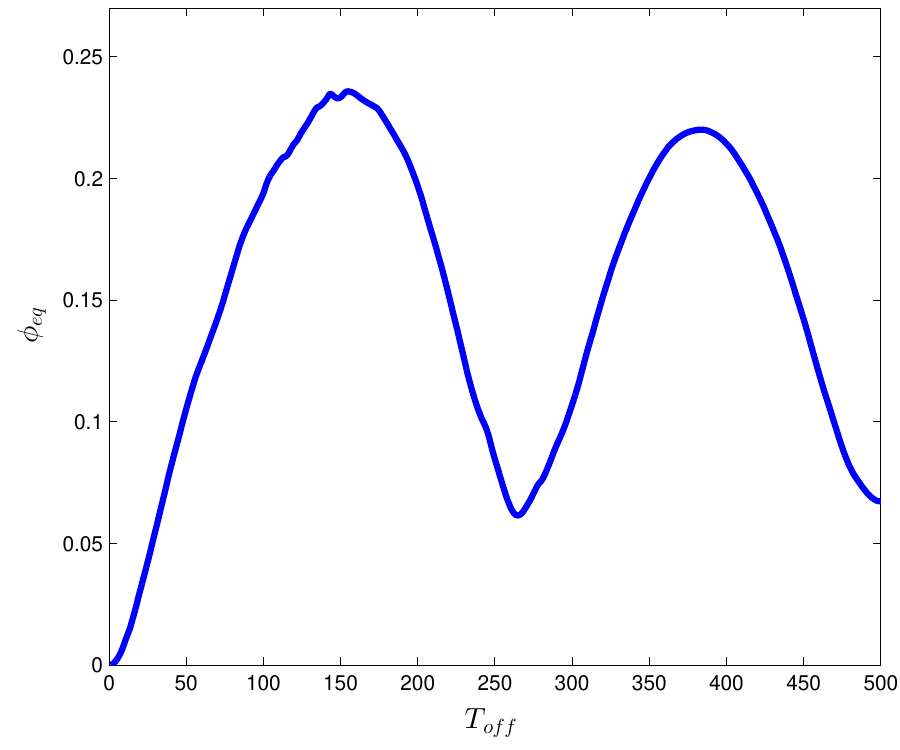}
\caption{(Color online) The amplitude $\phi_{eq}$ of  EPW as a function of $T_{off}$ for $\phi_{pump}=0.01$ and $k_z=0.35$.}
\label{fig:1D_phi_eq_vs_Toff}
\end{figure}

The resulting amplitude $\phi_{eq}$  depends on $\phi_{pump}$ and $T_{off}$. As we already discussed at the beginning of this Section, if we fix $\phi_{pump}$, there is only a certain range of amplitudes of EPW $0\leq\phi_{eq}\leq\phi_{eq}^{max}(\phi_{pump})$ that can be achieved by varying  $T_{off}$, where the dependence $\phi_{eq}^{max}(\phi_{pump})$ is obtained from simulations. To get $\phi_{eq}$ close to $\phi_{eq}^{max}$, we need to turn the pump off around (but not exactly) the time when  $\phi_1(t)$ is close to its global maximum as exemplified in Fig. \ref{fig:1D_phi_1_t}. To study this question more systematically we performed a series of simulations with $k_z=0.35$, $\phi_{pump}=0.01$ and various $T_{off}$ and obtained $\phi_{eq}$ as a function of $T_{off}$ (see Fig. \ref{fig:1D_phi_eq_vs_Toff}). The maximum $\phi_{eq}^{max}=0.2358$ is obtained if we choose $T_{off}\approx155$ while the global maximum of $\phi_1(t)$ is achieved at $t=t_{glob}=169$. This difference $t_{glob}-T_{off}$  is about one third of the bouncing period   $t_{glob}-T_{off}=14\simeq T_{bounce}/3\approx12$ estimated from $\phi_{eq}^{max}$.   Looking at other values $\phi_{pump}$ we found that typically the maximal value of $\phi_{eq}=\phi_{eq}^{max}$ can be obtained if the pump is switched off about $T_{bounce}/3$ before the global maximum of $\phi_1(t)$ is achieved. The same $\phi_{eq}$ can also be achieved by using a larger $\phi_{pump}$ (and respectively smaller $T_{off}$) but not a smaller $\phi_{pump}$. In this sense we obtain  EPW with amplitude $\phi_{eq}=\phi_{eq}^{max}$ using the smallest $\phi_{pump}$ possible (and correspondingly the largest $T_{off}$). We call these forcing parameters ($\phi_{pump}$ and $T_{off}$)  obtained for the given $\phi_{eq}=\phi_{eq}^{max}$ the {\it weak pumping}.
After such smallest $\phi_{pump}$ (together with $T_{off}$) is found for a given  $\phi_{eq}$ (or, in practice, we fix $\phi_{pump}$ and maximize $\phi_{eq}$ varying $T_{off}$), we run 2+2D simulations with the forcing \e{eq:PUMPPotential} as described above.

Tables  \ref{tab:Table1} and \ref{tab:Table2} provide a set of approximate values $\phi_{pump}$ and $T_{off}$
found by this procedure that we used for our 2+2D simulations for $k_z=0.35$ and $k_z=0.425$ correspondingly. We did not aim to obtain these  values with very high precision (but rather $\sim 20\%$ within the optimal values) because further increase in precision has a small effect on transverse instability growth rates. First three rows in Tables \ref{tab:Table1} and \ref{tab:Table2} are even more than $20\%$ away from optimal parameters $\phi_{pump}$ and $T_{off}$.

Another way to look at the degree of  ``strength" of the pumping of EPW is to see how many revolutions the spiral in the trapping region of the distribution function makes before the pumping is turned off. Following the estimates in Ref. \cite{BanksEtAlPhysPlasm2011}, we conventionally call the pumping {\it weak} if it makes more than one revolution during the pumping period or $\int_{0}^{T_{off}}{dt/T_{bounce}(t)}>1$, or equivalently $\int_{0}^{T_{off}}{\omega_{bounce}(t)dt}>2\pi$.  Assume that the pump is switched off not later than a global maximum of $\phi_1(t)$ is achieved and that $\phi_1(t)$ grows approximately linearly during $0<t<T_{off.}$ Also we estimate $\phi_{eq}$  as $\phi_{eq}\approx\phi_1(T_{off})$, then the condition of pumping strength  of Ref. \cite{BanksEtAlPhysPlasm2011} is reduced to $k_z \sqrt{\phi_{eq}} T_{off}>3\pi$ , i.e.
\begin{equation}
    T_{off}>3\pi/(k_z \sqrt{\phi_{eq}}).
    \label{eq:AdiabaticPUMP}
\end{equation}

All simulation parameters of  Tables \ref{tab:Table1} and \ref{tab:Table2}  satisfy the criterion of adiabaticity Eq. \e{eq:AdiabaticPUMP} except for the first row in Table \ref{tab:Table1} and the first three rows in Table \ref{tab:Table2}.

\begin{table}
\caption{Parameters of  simulations with weak pumping for $k_z=0.35, L_z=\frac {2\pi}{k_z}, v_z^{max}=8,v_x^{max}=6,$
$N_x=64, N_{v_x}=32$.
} \label{tab:Table1}
\begin{tabular}{|c|c|c|c|c|c|c|c|c|}
  \hline
  $\Delta t $ & $D_{16v_z}$ & $N_z$ & $N_{v_z}$ & $\phi_{pump}$ & $T_{off}$ & $\phi_{eq}$ & $T_{final}$ & $L_x$ \\
  \hline
  0.1 & $10^{-30}$ & 32  & 512 & 0.0005 & 200 & 0.007 & 20000 & 1600$\pi$ \\
  0.1 & $10^{-30}$ & 48  & 512 & 0.001  & 200 & 0.022 & 20000 & 1600$\pi$ \\
  0.1 & $10^{-25}$ & 48  & 256 & 0.002  & 200 & 0.053 & 10000 & 1600$\pi$ \\
  0.05& $10^{-25}$ & 48  & 256 & 0.003  & 210 & 0.085 & 7500  & 800$\pi$ \\
  0.05& $10^{-25}$ & 64  & 256 & 0.005  & 210 & 0.13  & 5000  & 800$\pi$ \\
  0.05& $10^{-25}$ & 64  & 256 & 0.01   & 110 & 0.20  & 4000  & 800$\pi$ \\
  0.05& $10^{-25}$ & 64  & 256 & 0.015  & 110 & 0.29  & 3000  & 400$\pi$ \\
  0.05& $10^{-25}$ & 96  & 256 & 0.02   & 120 & 0.38  & 3000  & 400$\pi$ \\
  0.05& $10^{-25}$ & 96  & 256 & 0.03   & 100 & 0.50  & 3000  & 400$\pi$ \\
  0.05& $10^{-25}$ & 96  & 256 & 0.04   & 100 & 0.59  & 2000  & 400$\pi$ \\
  0.05& $10^{-25}$ & 96  & 256 & 0.05   & 90  & 0.69  & 2000  & 400$\pi$ \\
  0.05& $10^{-25}$ & 128 & 256 & 0.06   & 80  & 0.77  & 2000  & 400$\pi$ \\
  0.05& $10^{-25}$ & 128 & 256 & 0.07   & 80  & 0.84  & 1500  & 400$\pi$ \\
  0.05& $10^{-25}$ & 128 & 256 & 0.1    & 70  & 1.01  & 1200  & 400$\pi$ \\
  \hline
\end{tabular}\\
\end{table}

\begin{table}
\caption{Parameters of  simulations with weak pumping for $k_z=0.425, L_z=\frac {2\pi}{k_z}, v_z^{max}=8,v_x^{max}=6,$
$N_x=64, N_{v_x}=32$.
} \label{tab:Table2}
\begin{tabular}{|c|c|c|c|c|c|c|c|c|}
  \hline
  $\Delta t $ & $D_{16v_z}$ & $N_z$ & $N_{v_z}$ & $\phi_{pump}$ & $T_{off}$ & $\phi_{eq}$ & $T_{final}$ & $L_x$ \\
  \hline
  0.1 & $10^{-30}$ & 64  & 512 & 0.002  & 100 & 0.0106& 7000 & 1600$\pi$ \\
  0.1 & $10^{-30}$ & 64  & 512 & 0.003  & 100 & 0.0195& 6000 & 1600$\pi$ \\
  0.1 & $10^{-30}$ & 64  & 512 & 0.005  & 100 & 0.036 & 6000 & 1600$\pi$ \\
  0.05& $10^{-25}$ & 48  & 256 & 0.007  & 100 & 0.052 & 5000  & 800$\pi$ \\
  0.05& $10^{-25}$ & 48  & 256 & 0.01   & 100 & 0.075 & 5000  & 800$\pi$ \\
  0.05& $10^{-25}$ & 48  & 256 & 0.016  & 60  & 0.10  & 3500  & 800$\pi$ \\
  0.05& $10^{-25}$ & 64  & 256 & 0.025  & 60  & 0.15  & 2500  & 400$\pi$ \\
  0.05& $10^{-25}$ & 64  & 256 & 0.035  & 60  & 0.21  & 2000  & 400$\pi$ \\
  0.05& $10^{-25}$ & 96  & 256 & 0.06   & 50  & 0.31  & 1600  & 400$\pi$ \\
  0.05& $10^{-25}$ & 128 & 256 & 0.13   & 35  & 0.51  & 1100  & 400$\pi$ \\
  0.05& $10^{-25}$ & 128 & 256 & 0.2    & 30  & 0.63  & 1000  & 200$\pi$ \\
  0.05& $10^{-25}$ & 128 & 256 & 0.25   & 30  & 0.73  & 8000  & 200$\pi$ \\
  0.05& $10^{-25}$ & 128 & 256 & 0.4    & 27  & 0.86  &  600  & 200$\pi$ \\
  \hline
\end{tabular}\\
\end{table}

\subsection{Creation of BGK-like solutions via strong external pumping}
\label{sec:FastEPW}
After the weak pump parameters ($\phi_{pump}$ and $T_{off}$) are found for the desired amplitude of EPW $\phi_{eq}$ we can find stronger pumping parameters (with larger $\phi_{pump}$ and smaller $T_{off}$) that provide the same $\phi_{eq}$. Typically, if we want to keep $\phi_{eq}$ fixed and increase $\phi_{pump}$ two times we need to decrease $T_{off}$ a little more than two times. In the limit $\phi_{pump}\rightarrow\infty$ and $T_{off}\rightarrow0,$ the action of the pump becomes equivalent to an initial perturbation of electron density in $z$-direction by $\delta$-function in time followed by a further evolution of the system without external pump.

To study the difference in terms of transverse instability of EPWs obtained by weak and strong pumps we performed another set of simulations with parameters like in Table  \ref{tab:Table1} with the only difference that $\phi_{pump}$ was 10 times larger and $T_{off}$ was 10 times smaller than in Table \ref{tab:Table1}. We  call such pumping parameters by {\it strong pumping}. The corresponding amplitudes $\phi_{eq}$ for the  strong pumps were $30-60\%$ larger than for the weak pumps. They could have been matched to the amplitudes $\phi_{eq}$ of corresponding weak pumps by further adjusting $T_{off}$, but it was not necessary for us below since we were comparing the simulations not one-to-one but rather a set of simulations with weak pumps to a set of simulations with strong pumps.
All the simulations with strong pumping do not satisfy the criterion of adiabaticity  Eq. \e{eq:AdiabaticPUMP}.

The nonlinear frequency shift of the resulting EPW depends on the way it was created. Two limiting cases for finite amplitude EPWs have been
treated analytically by Dewar in Ref. \cite{DewarPhysFL1972} providing the nonlinear frequency shift approximation
\begin{eqnarray}
\label{eq:dW_Dewar}
  \Delta \omega_{NL}^{Dewar} = -\alpha \left[ \frac{ \p\varepsilon_0(\omega_{LW}) }{\p\omega} \right]^{-1} f_0''(v_\varphi) \frac{\sqrt{\phi_{eq}}}{k_z^2},
\end{eqnarray}
where $\varepsilon_0$ is linear dielectric function given by Eq. (23) in Part I \cite{SilantyevLushnikovRosePOP2017} and $\alpha = 0.77\sqrt{2}=1.089$ and $\alpha=1.163\sqrt{2}=1.645$ for the ``adiabatic" and ``sudden" excitation of nonlinear LW, respectively. Our {\it weak} pump is only somewhat adiabatic in Dewar's sense since its amplitude stays constant for the whole time of driving EPW rather than slowly varying. Our {\it strong} pump is closer to the sudden case in Dewar's theory yet still no exactly the same since after turning off the external pump our EPW still evolves while Dewar considers the asymptotic limit in which the distribution function is constant along the lines of constant wave-frame energy. $\frac{ \p\varepsilon_0(\omega_{LW}) }{\p\omega}=2.267$ for $k_z=0.35$ and $\frac{ \p\varepsilon_0(\omega_{LW}) }{\p\omega}=1.781$ for $k_z=0.425$.

\section{Transverse instability of BGK-like solution}
\label{sec:TransverseInstability}

After the pumping is turned off at $t=T_{off}$,  BGK-like solution with the amplitude $\phi_{eq}$ continue to slowly evolve as shown in Fig. \ref{fig:1D_phi_1_t} and described in Section \ref{sec:AdiabaticEPW}. During that slow evolution, the transverse instability of BGK-like solution starts to develop. We look at the linear stage of that instability analytically through the solution in the moving frame in the following form
\begin{equation}
    \Phi =\Real\left\{\exp(ik_z z)[\phi_{eq}+ \delta\phi(t)\exp(i{\bf\delta k \cdot r}) ]\right \},
    \label{eq:Anzac}
\end{equation}
where the wave vector $\delta \bfk \perp \hat z$ is responsible for the transverse perturbations with the amplitude $\delta\phi(t)
$. Here $\hat z$ is the unit vector in $z$ direction. Let $\delta\phi \sim exp(\gamma t)$. Assuming that $\phi_{eq} $ does not change with time, we use the result of  Ref. \cite{RoseYinPOP2008}  outlined in Part I, that
\begin{equation}
    (\gamma + \nu_{residual})^2 =-D\left (\phi_{eq} \frac{\partial\omega}{\partial\phi_{eq}} +D\right),
    \label{eq:GrowtRate}
\end{equation}
where $D$ is  the generalized diffraction operator given by
\begin{equation}
    D =\omega(|k_z\hat z+\delta \textbf{k}|,\phi_{eq})-\omega(k_z,\phi_{eq}),
    \label{eq:D2}
\end{equation}
and  $\omega(k_z,\phi_{eq})$ is the nonlinear frequency of BGK-like solution with the amplitude $\phi_{eq}$. Contrary to Part I, we recover that frequency directly from simulations as the rate of change of phase. 

Additionally, assuming  $\phi_{eq} \ll 1,$  we
 approximate $\omega(k_z,\phi_{eq}), \ \phi_{eq}\to 0$  through the liner LW dispersion relation $\omega_{LW}(k_z)$ (obtained using $Z$-function \cite{FriedGell-MannJacksonWyldJNF1960}, see e.g. Refs. \cite{PitaevskiiLifshitzPhysicalKineticsBook1981,NicholsonBook1983}). Also assuming  $|\delta{\bf k}|\ll 1$,     we reduce Eq. \e{eq:D2}  to the  following expression
\begin{align} \label{eq:Dlin}
    D \approx D_{lin}=\left .\frac{1}{2k_z} \frac{\p \omega_{LW}(|\textbf{k}|) }{\p |\textbf{k}|} \right|_{|\textbf{k}|=k_z}|\delta\textbf{k}|^2\nonumber \\=\frac{v_g^{LW}}{2k_z} |\delta\textbf{k}|^2, \quad  v_g^{LW}\equiv\p \omega_{LW}(k_z)/\p k_z,
\end{align}
where $v_g^{LW}  $ is the linear LW group velocity. Also the residual damping, $\nu_{residual}$, from Eq. (\ref{eq:GrowtRate}) is model dependent and, as we discussed in Part I, we  set $ \nu_{residual}=0$ in \e{eq:GrowtRate} as it is the only choice that appears to be consistent with our simulations.

For the term $\phi_{eq} \frac{\partial\omega}{\partial\phi_{eq}}$ in Eq. (\ref{eq:GrowtRate}), we have to take into account the dependence on $\phi_{eq}$. Assuming at the leading order that the nonlinear frequency shift $\Delta\omega\equiv\omega(k_z,\phi_{eq})-\omega_{LW}(k_z)\varpropto\sqrt{\phi_{eq}}$ we obtain that  $\phi_{eq} \frac{\partial\omega}{\partial\phi_{eq}}=\Delta\omega/2.$ Maximizing $\gamma$ over $D$ in Eq. \e{eq:GrowtRate}
we get the maximum value
\begin{equation} \label{gammamaxsqrt}
\gamma^{max}=|\Delta\omega|/4,
\end{equation}
at
\begin{equation} \label{Dmax}\
D=-\Delta\omega/4,
\end{equation}
which is valid for $|\bf{\delta k}| \ll |k|$.
Using the approximation \e{eq:Dlin}, we obtain from Eq. \e{Dmax} the position of the maximum
\begin{equation} \label{kmaxapprox}
|{\bf{\delta k}}|= k_x^{max}=\left ( \frac{-\Delta\omega \,k_z}{2v_g^{LW}   }\right )^{1/2}.
\end{equation}
%

\section{2+2D simulations and instability of BGK-like EPWs}
\label{sec:2DPumpedEPW}
We performed two types of 2+2D fully non-linear Vlasov simulations to study the transverse instability of nonlinear electron plasma waves that are dynamically prepared by starting with uniform in space initial conditions with Maxwellian distribution of particle velocities and pumping the system by both weak and strong pumps described in  Section \ref{sec:1DEPW}.

\subsection{2+2D Simulation settings and methods}
\label{sec:Simulationsettings}

In both cases we simulate 2+2D Vlasov-Poisson system  (\ref{eq:vlasov})-(\ref{eq:density}) in the phase space $(x,z,v_x,v_z)$  using fully spectral (i.e. spectral in all four dimensions) code and split-step (operator splitting) method of 2nd order in time with periodic boundary conditions (BC) in all four dimensions. To ensure a spectral convergence and imitate the weak effect of collisions, we added to Eq. (\ref{eq:vlasov}) a small additional hyper-viscosity term as follows:

\begin{equation}
\begin{split}
     &\left\lbrace\frac{\partial }{\partial t} + v_z\frac{\partial }{\partial z} + v_x\frac{\partial }{\partial x} + E_z \frac{\partial }{\partial v_z} + E_x \frac{\partial }{\partial v_x} \right\rbrace f= \\  
     &- D_{16v_z}\frac{\p^{16}}{\p v_z^{16}}\left(f - \frac{1}{L_z}\int_0^{L_z} f dz\right),
    \label{eq:2DVlasov}
\end{split}
\end{equation}
where  $D_{16v_z}$ is the 16th order hyper-viscosity coefficient. The hyper-viscosity term in the right-hand side (r.h.s.) of Eq. \e{eq:2DVlasov} is used to prevent recurrence \cite{ChengKnorrJCompPhys1976} and aliasing (which causes propagation of numerical error from high modes to low modes) effects. We use periodic BC in $z$ direction with the period $L_z=2\pi/k_z$. Choosing   $L_z=2\pi/k_z $ allows to focus on the study of transverse instability effects (along $x$) while avoid subharmonic (sideband instability) \cite{KruerDawsonSudanPRL1969} in longitudinal $z$-direction.  Periodic BC in $x$ with the period $L_x$ together with  $x$-independent initial condition (IC) are used to separate transverse instability effects from any sideloss effects due to trapped electrons traveling in the transverse direction (this is in contrast to Ref. \cite{LushnikovRoseSilantyevVladimirovaPhysPlasmas2014}, where the transverse spatial profile in the initial condition made sideloss comparable with the transverse growth rate).  We chose typically $200\pi\le L_x\le 1600\pi$ depending on amplitude of EPW to capture all growing transverse modes. The rest of the simulation settings are provided  in Part I.

\subsection{2+2D simulations and transverse instability of nonlinear EPWs}
\label{sec:transverseBGKlike}
\begin{figure}
\includegraphics[width=3in]{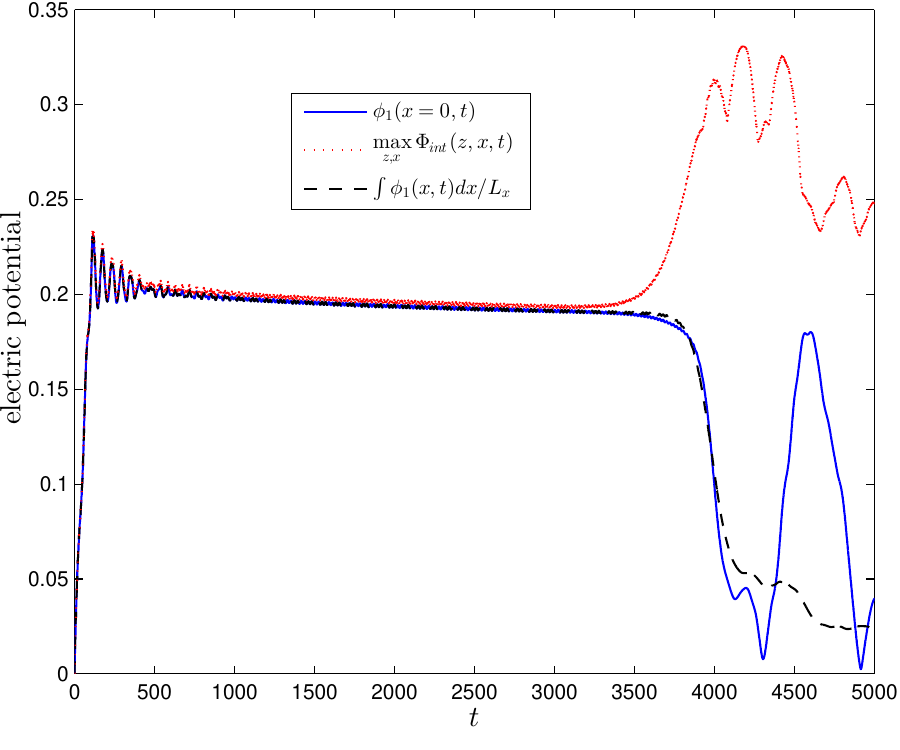}
\caption{(Color online) Solid line is for the first $z-$harmonic $\phi_1(x,t)$ evaluated at $x=0$, dashed line is for the averaged value  $\int^{L_x}_0\phi_1(x,t)dx/L_x$  and dotted line is for the maximum of electrostatic potential $\max\limits_{z,x}\Phi_{int}(z,x,t). $ Simulation parameters are $\phi_{pump}=0.01, T_{off}=110$ and $k_z=0.35$. }
\label{fig:Phi_int}
\end{figure}

\begin{figure}
\includegraphics[width=3in]{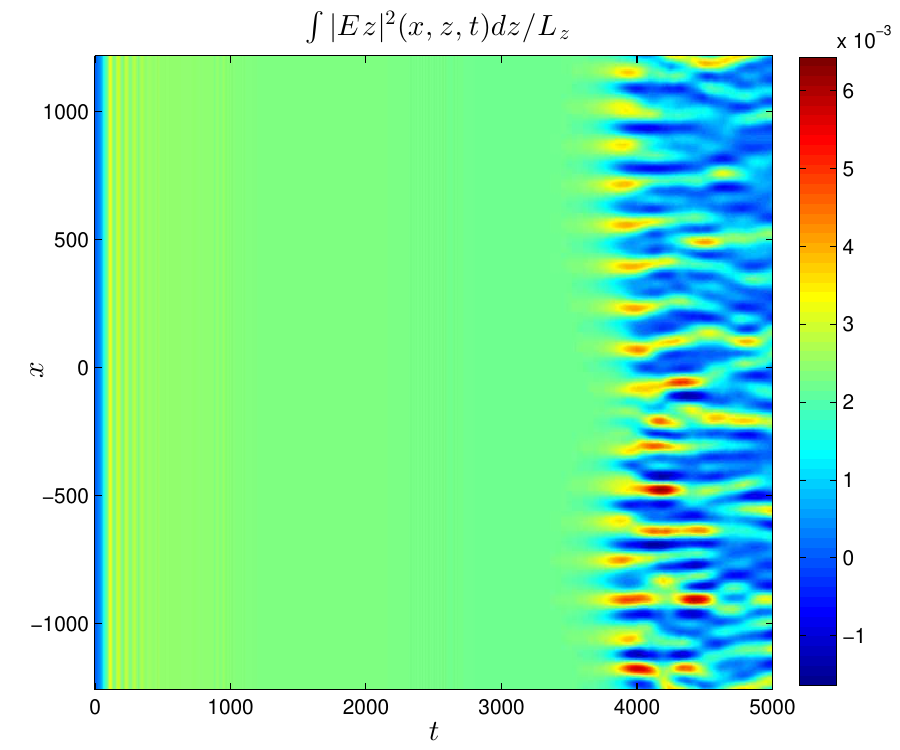}
\caption{(Color online) The density plot vs.  $x$ and $t$ for   $\langle|E_z|^2\rangle_z\equiv \int^{L_z}_0|E_z|^2dz/L_z$ ($|E_z|^2$
averaged over $z$) shows a development of LW filamentation with time from the pumped EPW with $k_z=0.35, \phi_{eq}\approx 0.2$ . }
\label{fig:Filamentation_EPW}
\end{figure}

\begin{figure}
\includegraphics[width=3in]{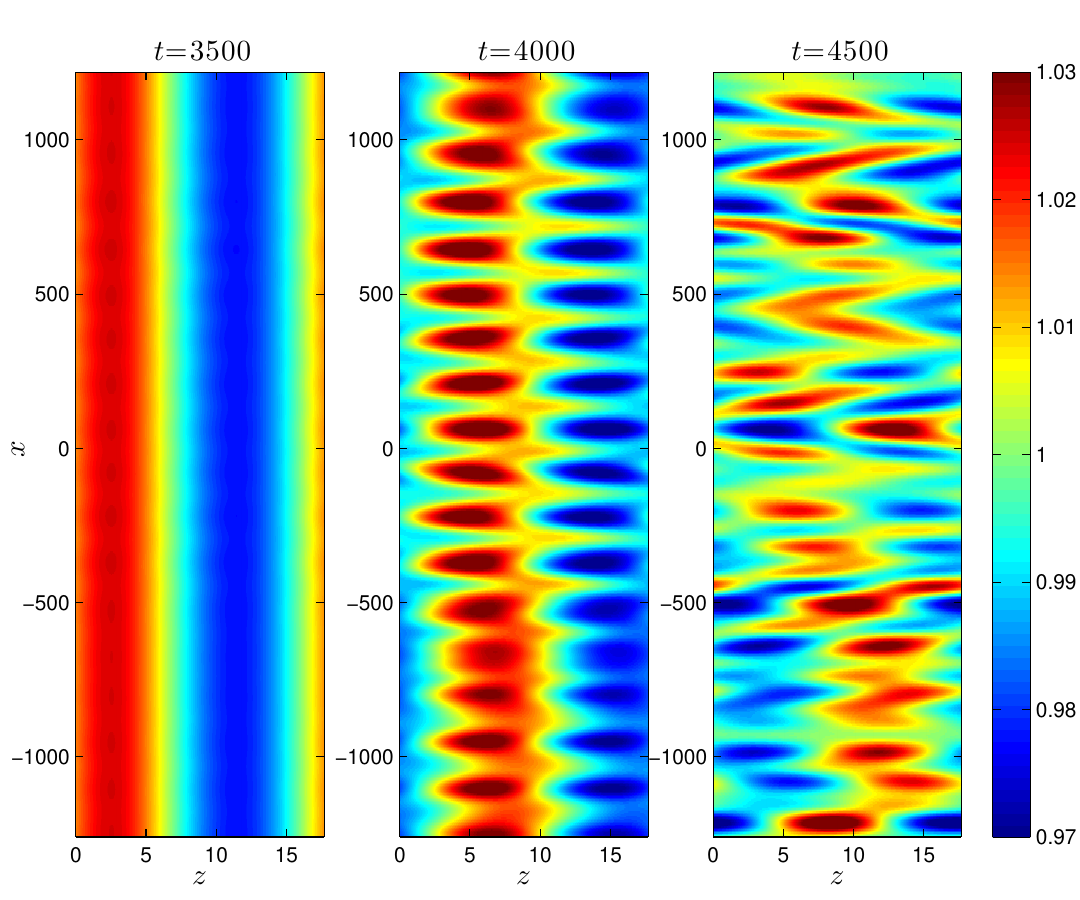}
\caption{(Color online) Density plot of particle density $\rho(z,x)$  before ($t=3500$), during ($t=4000$) and after ($t=4500$) LW filamentation for pumped EPW with $k_z=0.35, \ \phi_{eq}\approx 0.2$.}
\label{fig:RhoModulation_EPW}
\end{figure}

We start by presenting an example of a simulation with $k_z=0.35$, $\phi_{pump}=0.01, \ T_{off}=110$ and resulting $\phi_{eq}\approx0.2$. Fig. \ref{fig:Phi_int} shows the amplitude of the electrostatic potential  $\Phi_{int}(z,x,t)$ vs. $t$. Solid line is for the first $z-$harmonic, $\phi_1(x,t)\equiv 2|\int^{L_z}_0 \Phi_{int}(z,x,t)\exp{(ik_zz)}dz|/L_z $ evaluated at $x=0$, dashed line is for the averaged value  $\langle \phi_1\rangle_x= \int^{L_x}_0\phi_1(x,t)dx/L_x$  and dotted line is for the maximum of electrostatic potential $\max\limits_{z,x}\Phi_{int}(z,x,t)$. Other simulation parameters were $D_{16v_z}=10^{-25}$, $64\times256\times64\times32$ grid points for $(z,v_z,x,v_x)$ with $L_z=2\pi/k_z, L_x=800\pi, v_z^{max}=8,v_x^{max}=6, \Delta t=0.05, T_{final}=5000$. It is seen in Fig.  \ref{fig:Phi_int} that during the action of pumping $\langle \phi_1\rangle_x$  reaches the global maximum. Then after pumping is switched off, $\langle \phi_1\rangle_x$ experiences a short initial transient behaviour, after that it   remains almost constant until $t\sim 3500$, after that a strong LW filamentation occurs at  $t\sim4000$ (see Figs. \ref{fig:Filamentation_EPW} and \ref{fig:RhoModulation_EPW}). During the long quasi-stationary dynamics $500\lesssim t\lesssim 3500,$ we call  the  quasi-equilibrium value of  $\langle \phi_1\rangle_x$ by  $\phi_{eq}$. In Fig.  \ref{fig:Phi_int}  $\phi_{eq}\approx0.2$. LW filamentation peaks after $t=4000$  with the value of $\max\limits_{z,x}\Phi_{int}(z,x,t)$ almost twice higher than before filamentation. At that time, a large portion of electrostatic field energy from the first Fourier mode (that has the most of electric field energy) is transferred into kinetic energy as can be seen from the dynamics of $\langle \phi_1\rangle_x$.


\begin{figure}
\includegraphics[width=3in]{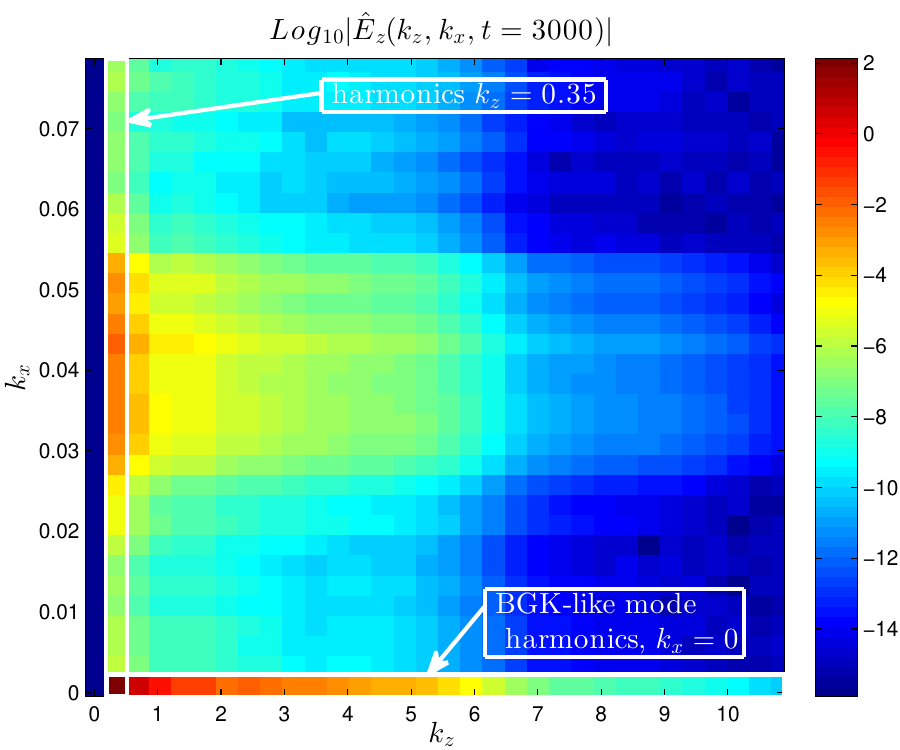}
\caption{(Color online) The density plot of the spectrum of $E_z(z,x)$ at  $t=3000$.}
\label{fig:EzSpectrum}
\end{figure}

\begin{figure}
\includegraphics[width=3in]{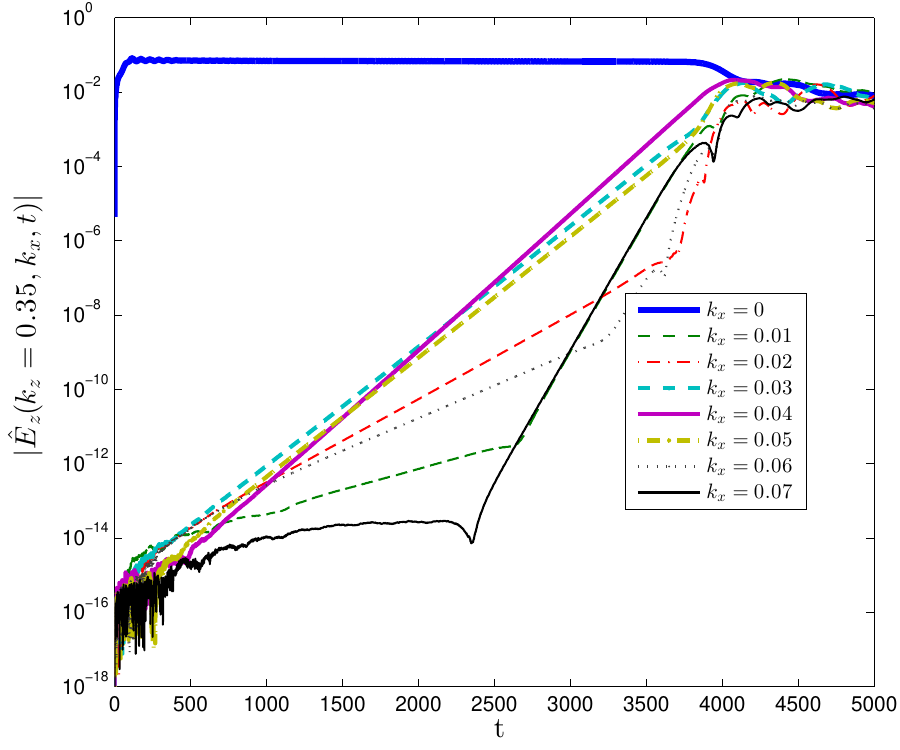}
\caption{(Color online) The growth of harmonics $|\hat E_z(k_z=0.35,k_x,t)|$ in time.}
\label{fig:EzGrowth}
\end{figure}

\begin{figure}
\includegraphics[width=3in]{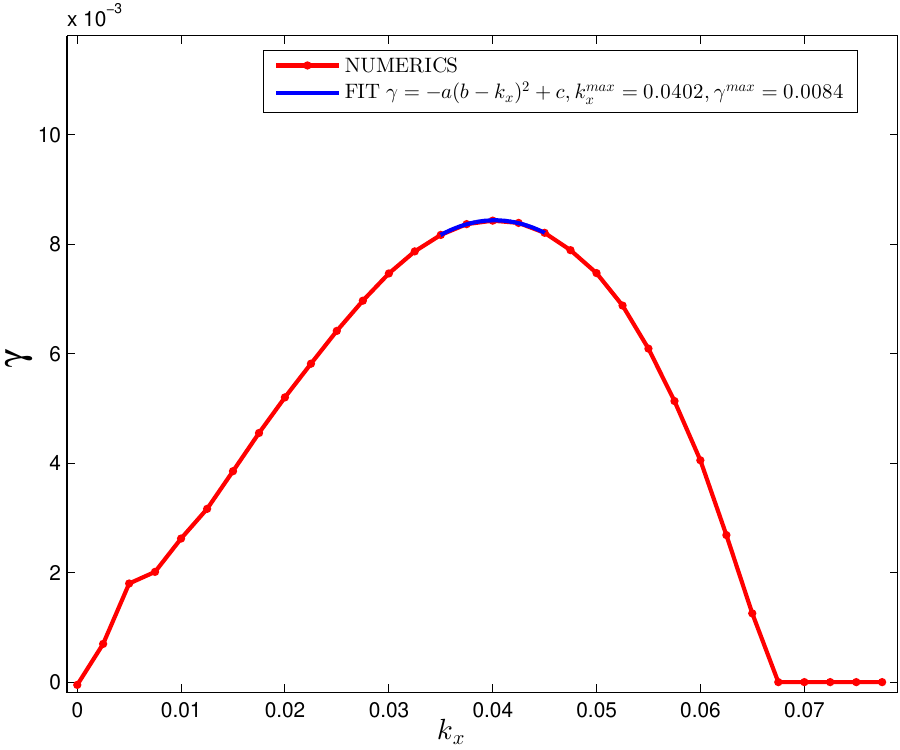}
\caption{(Color online) The growth rates $\gamma_{k_x}$ of oblique harmonics extracted from the least-square fit of the data of Fig. \ref{fig:EzGrowth}. Also shown a fit to the quadratic law near the maximum.}
\label{fig:G(kx)}
\end{figure}

We run the simulation for a long enough time (after the pumping is off) to observe the growth of oblique harmonics of electric field with wave vectors $(k_z=0.35,k_x)$ (see Fig. \ref{fig:EzSpectrum} for the spectrum of $E_z$, the $z$ component of the electric field) in several orders in magnitude (see Fig. \ref{fig:EzGrowth}), where $k_z$ is the wavenumber of the pump and $k_x$ varies between $-k_x^{max}$ and $k_x^{max}=\pi/\Delta x$. Here $\Delta x=L_x/N_x$, where $N_x$ is the number of grid points in $x$. The initial values in these harmonics are near the machine precision. During the simulation they grow from values $\sim10^{-16}$ to $\sim10^{-1}$. The exponential growth rates $\gamma_{k_x}$ for these harmonics are extracted (see Fig. \ref{fig:G(kx)}) when amplitudes grow from $\sim10^{-13}$ to $\sim10^{-8}-10^{-6}$ (during these times a clear exponential growth $\propto \exp{(\gamma_{k_x}t)}$ is observed before the nonlinear effects become noticeable). In Fig. \ref{fig:G(kx)} the maximum growth rate $\gamma^{max}$ (the maximum over $k_x$ for each fixed  $\phi_{eq}$) and $k_x^{max}$ are found using quadratic fit to several data points around the maximum.


These kind of simulations were done for a variety of pumped EPWs with $k_z=0.35$ and $k_z=0.425$  and amplitudes $0.007 \leq \phi_{eq} \leq 1.$ For $k_z=0.35$ we also considered two cases of pumping (weak and strong) as described in Section \ref{sec:1DEPW}. Parameters typically used for simulations were $D_{16v_z}=10^{-30}-10^{-25}$, the time step $\Delta t=0.05-0.1$, the final simulation time $T_{final}$ in the range $ 1000\le T_{final}\le 20000$ (depending on EPW amplitude $\phi_{eq}$) and from $32\times 256\times 64\times 32 $ up to $128\times 512\times 64\times 32 $ grid points for $(z,v_z,x,v_x)$ with $L_z=2\pi/k_z, L_x=200\pi-1600\pi, v_z^{max}=8,v_x^{max}=6$. Smaller amplitudes waves have narrower trapping region which requires more grid points and smaller hyper-viscosity coefficient to keep errors at approximately the same level in all of the of simulations. All parameters for simulations with weak pumping and $k_z=0.35$ are collected in Table \ref{tab:Table1} and with $k_z=0.425$ are collected in Table \ref{tab:Table2}. The simulations with strong pumping and $k_z=0.35$ were done with the same parameters as in Table \ref{tab:Table1} with the only difference that $\phi_{pump}$ was 10 times larger and $T_{off}$ was 10 times smaller.

\begin{figure}
\includegraphics[width=3in]{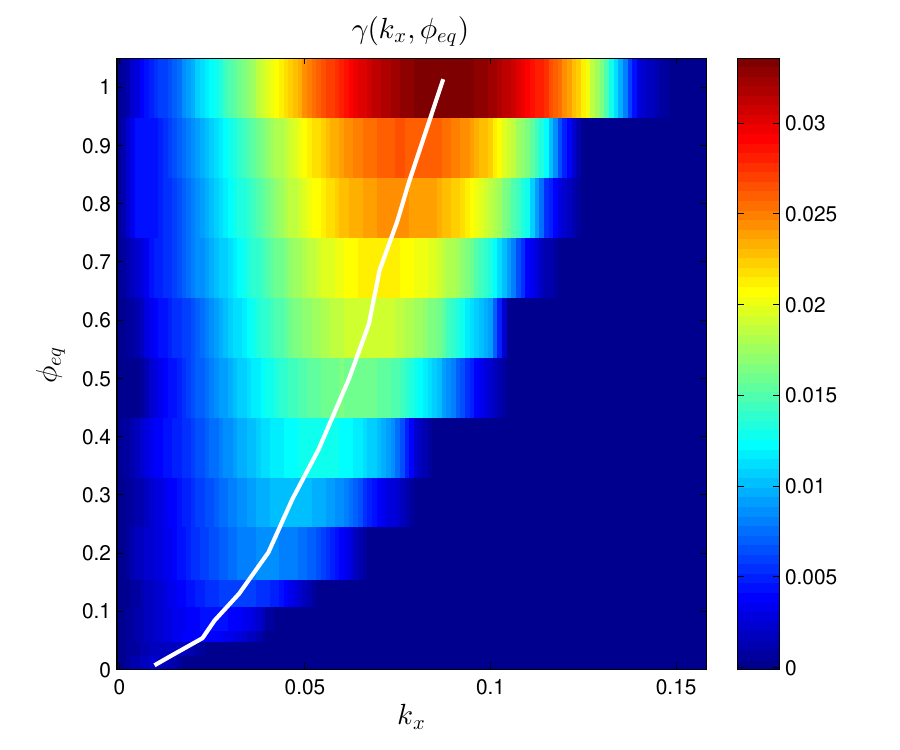}
\caption{(Color online) The density plot of the growth rate $\gamma_{k_x}$  as a function of $k_x$ and $\phi_{eq}$ for $k_z=0.35$, $v_{\varphi}=3.488$. The white line shows the maximum  $\gamma_{k_x}$ over $k_x$ for each $\phi_{eq}.$        }
\label{fig:G_vs_kx_vs_phi_EPW}
\end{figure}

\begin{figure}
\includegraphics[width=3in]{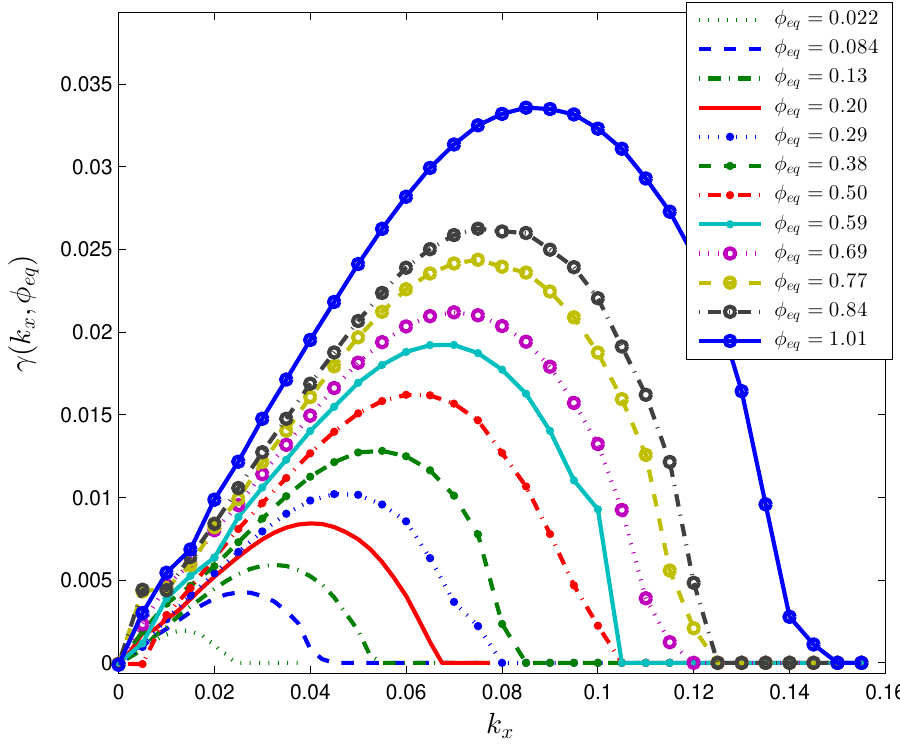}
\caption{(Color online) The growth rates  $\gamma_{k_x}$   as a function of $k_x$ for EPWs with various amplitudes $\phi_{eq}$ correspond to  multiple cross-sections of Fig. \ref{fig:G_vs_kx_vs_phi_EPW}. $k_z=0.35$, $v_{\varphi}=3.488$.}
\label{fig:G_vs_kx_vs_phi_EPW_}
\end{figure}

Figs. \ref{fig:G_vs_kx_vs_phi_EPW} and \ref{fig:G_vs_kx_vs_phi_EPW_} show the measured growth rates as a function of $k_x$ and $\phi_{eq}$ obtained from a set of simulations with $k_z=0.35$ and weak pump. We can clearly see the transverse instability for the whole range of amplitudes with higher amplitudes yielding larger growth rates.


In the further discussion we also overlayed data from Refs. \cite{BergerBrunnerBanksCohenWinjumPOP2015,BanksPrivate2016} that were produced in somewhat similar way (by pumping the system with external electric field in longitudinal direction for $0<t<100=T_{off}$ and measuring growth rates afterwards, however without systematic attempts to minimize  $\phi_{pump}$) using different kind of numerical scheme and turning on and off external pumping smoothly with $\tanh(t)$ function. Comparing smooth and non-smooth ways of turning the pump on and off in our simulations, we observed that the differences in results were negligible. Also we used $k_z=0.35$ and corresponding $\omega_{LW}(kz=0.35)=1.22095\ldots$ and $v_{\varphi}=3.488$ in our first set of simulations while  Refs. \cite{BergerBrunnerBanksCohenWinjumPOP2015,BanksPrivate2016}  used $k_z=1/3$ and corresponding $\omega_{LW}(k_z=1/3)=1.2$ and $v_{\varphi}=3.6$ which accounts for $5\%$ difference in $k_z$, $1.7\%$ difference in $\omega_{LW}$ and $3.2\%$ difference in $v_{\varphi}$, but we overlayed these data on the same graphs anyways for comparison. Second set of simulations was performed with exactly matching parameters $k_z=0.425$, $\omega_{LW}(k_z=0.425)=1.3176$ and $v_{\varphi}=3.1$ for both our simulations and simulations from Refs. \cite{BergerBrunnerBanksCohenWinjumPOP2015,BanksPrivate2016}.

\begin{figure}
\includegraphics[width=2.98in]{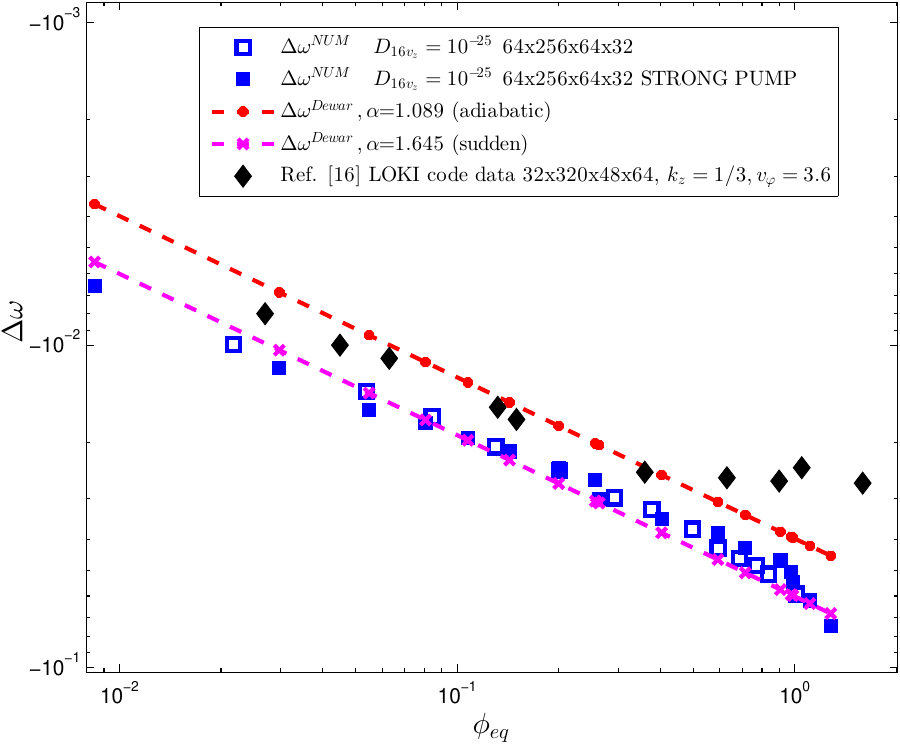}
\caption{(Color online) The nonlinear frequency shift $\Delta\omega$ as a function of $\phi_{eq}$ for $k_z=0.35$, $v_{\varphi}=3.488$.}
\label{fig:delta_W_vs_phi_EPW}
\end{figure}

During the simulations we extract the nonlinear frequency shift $\Delta\omega^{NUM}$ from simulations by finding the wave frequency as the rate of change of the phase of $\phi_1(x=0,t)$ and subtracting the reference value $\omega_{LW}(k_z)$. Fig. \ref{fig:delta_W_vs_phi_EPW} shows the nonlinear frequency shift $\Delta\omega^{NUM}$ for both weak and strong pumping (denoted as ``STRONG PUMP" in the legend) obtained from simulations in comparison with theoretical ones computed using Dewar's \cite{DewarPhysFL1972} nonlinear frequency shift approximation as in Eq. \e{eq:dW_Dewar} for the cases of adiabatic ($\alpha =  1.089$) and sudden ($\alpha=1.645$) excitations. The measured nonlinear frequency shift  $\Delta\omega^{NUM}$ is nearly the same for both weak and strong pumping and is close to $\Delta\omega^{Dewar}$ with $\alpha=1.645$. Also $\Delta\omega^{NUM}$  is mostly within Dewar's bounds (with $\alpha =  1.089$ and $\alpha=1.645$) and scales as $\Delta \omega \varpropto \sqrt{\phi_{eq}}$ for the whole range of amplitudes. Also we overlayed the data from Refs.  \cite{BergerBrunnerBanksCohenWinjumPOP2015,BanksPrivate2016} for comparison. It shows $\sim 30\%$ smaller nonlinear frequency shift since it was produced for $k_z=1/3$, $v_{\varphi}=3.6$ and exhibits different scaling for $\phi_{eq}>0.4$. If we were to plot the corresponding Dewar's bounds for the parameters $k_z=1/3$, $v_{\varphi}=3.6$ we would see that their nonlinear frequency shift data are also within those bounds for $\phi_{eq}<0.4$.

\begin{figure}
\includegraphics[width=2.98in]{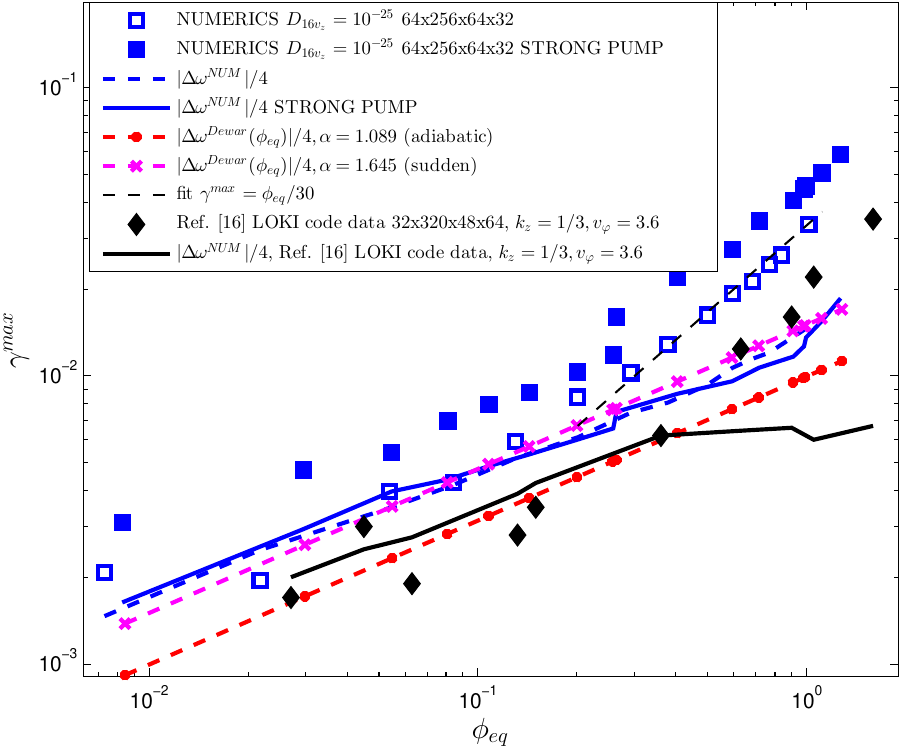}
\caption{(Color online) The maximum growth rate as a function of $\phi_{eq}$ for $k_z=0.35$, $v_{\varphi}=3.488$.}
\label{fig:G_max_vs_phi_EPW}
\end{figure}

The maximum growth rate $\gamma^{max}$ (the maximum over $k_x$ for each fixed  $\phi_{eq}$) as a function of $\phi_{eq}$ is shown in Fig. \ref{fig:G_max_vs_phi_EPW} together with the perturbative theoretical predictions given by Eq. \e{gammamaxsqrt} with different estimates for $\Delta\omega$ including Dewar's model \e{eq:dW_Dewar} and  $\Delta\omega^{NUM}$ recovered directly from simulations (with weak and strong pumps, respectively). We see that theoretical prediction $\gamma^{max}\approx|\Delta\omega^{NUM}|/4$ from Eq. \e{gammamaxsqrt} works pretty well for EPWs obtained with weak pump and $\phi_{eq}<0.2.$ In this case the measured growth rates are within $20-25\%$ from the estimate, and scale like $\gamma^{max} \varpropto \sqrt{\phi_{eq}}$. The measured growth rates for the strong pump are $30-50\%$ larger compared to the weak pump growth rates and also larger than a corresponding estimate $|\Delta\omega^{NUM}|/4$ in the whole range of amplitudes $\phi_{eq}$. Also for amplitudes $\phi_{eq}>0.3,$ the scaling changes for both weak and strong pumps and becomes $\gamma^{max} \varpropto \phi_{eq}$. The data from Refs. \cite{BergerBrunnerBanksCohenWinjumPOP2015,BanksPrivate2016} exhibit similar behaviour regarding the scalings and match the corresponding  estimate $\gamma^{max}\approx|\Delta\omega^{NUM}|/4$ for amplitudes $\phi_{eq}<0.4$.

\begin{figure}
\includegraphics[width=2.98in]{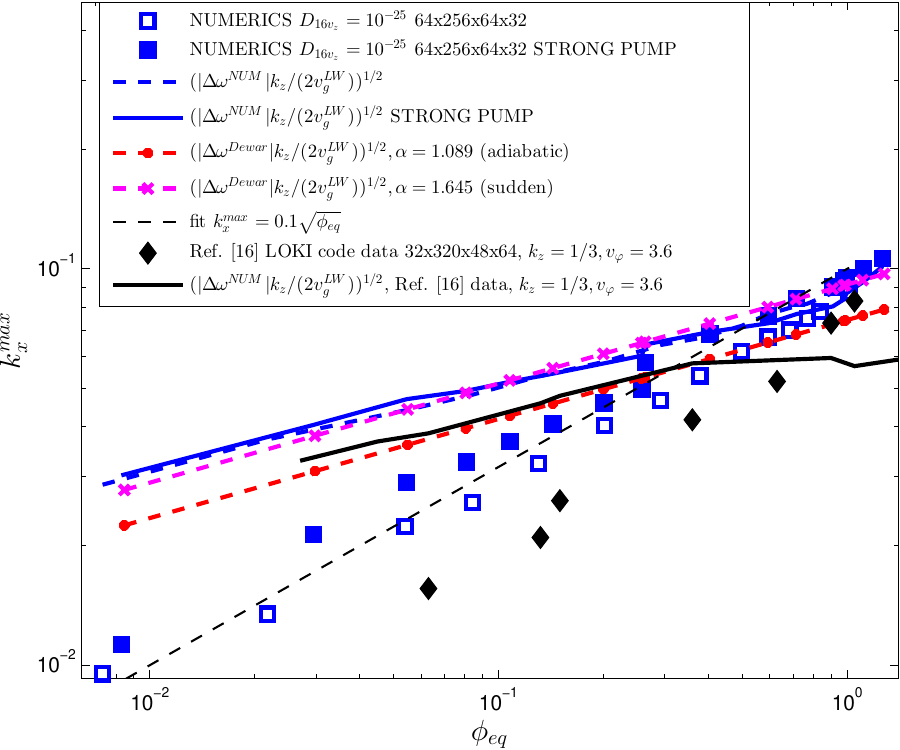}
\caption{(Color online) The wavenumber $k_x^{max}$ at which the growth rate reaches the maximum as a function of $\phi_{eq}$ for $k_z=0.35$, $v_{\varphi}=3.488$.}
\label{fig:kx_max_vs_phi_EPW}
\end{figure}

\begin{figure}
\includegraphics[width=2.98in]{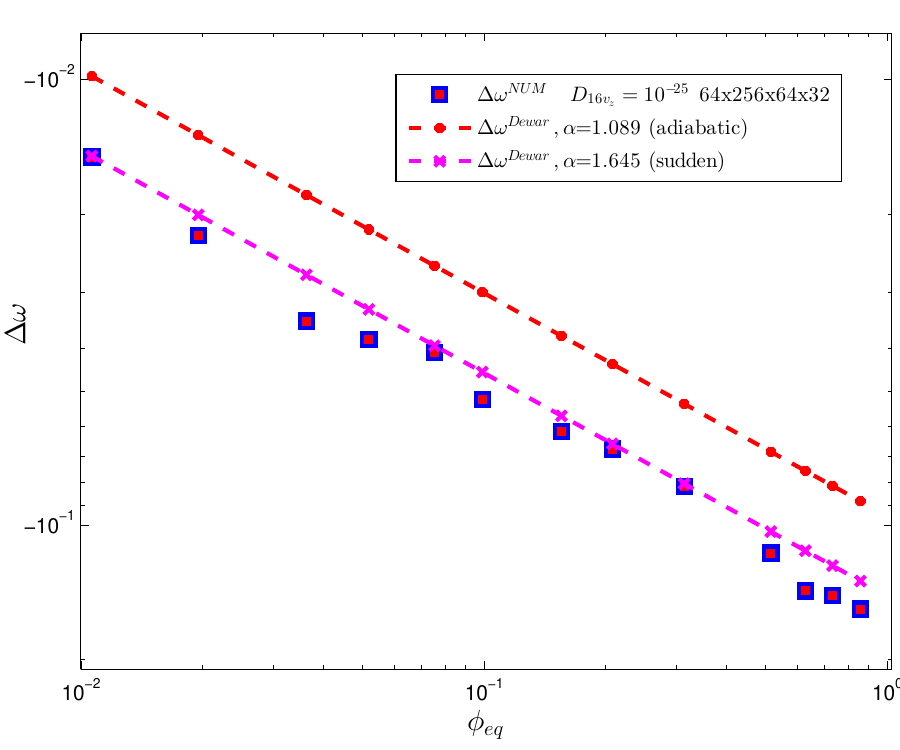}
\caption{(Color online) The nonlinear frequency shift  $\Delta\omega$ as a function of $\phi_{eq}$ for $k_z=0.425$, $v_{\varphi}=3.1$.}
\label{fig:delta_W_vs_phi_EPW_0425}
\end{figure}
\begin{figure}
\includegraphics[width=2.98in]{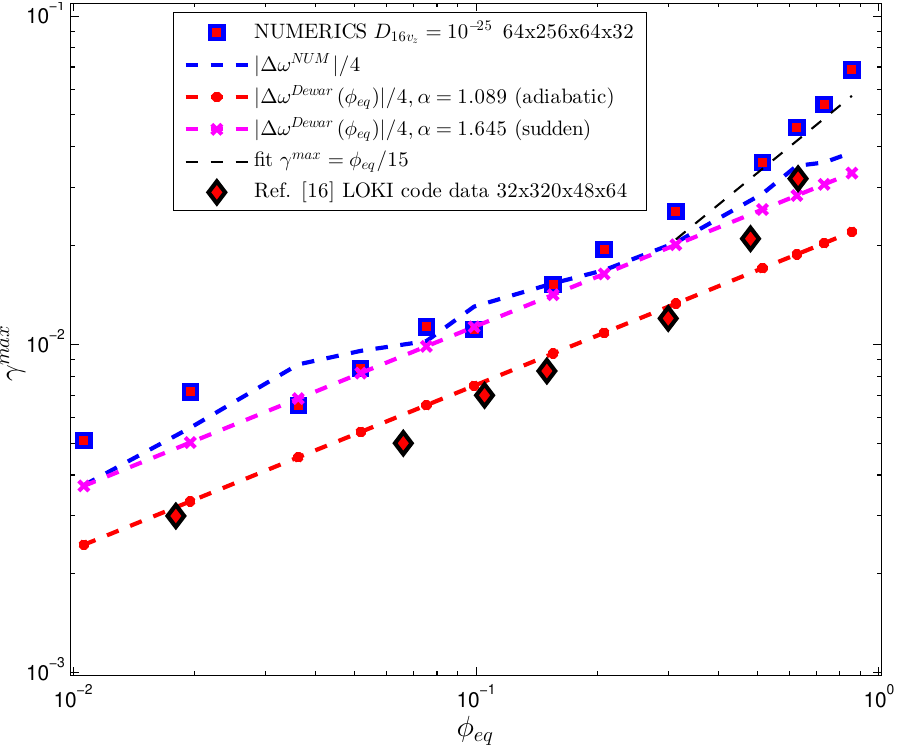}
\caption{(Color online) The maximum growth rate as a function of $\phi_{eq}$ for $k_z=0.425$, $v_{\varphi}=3.1$.}
\label{fig:G_max_vs_phi_EPW_0425}
\end{figure}
\begin{figure}
\includegraphics[width=2.98in]{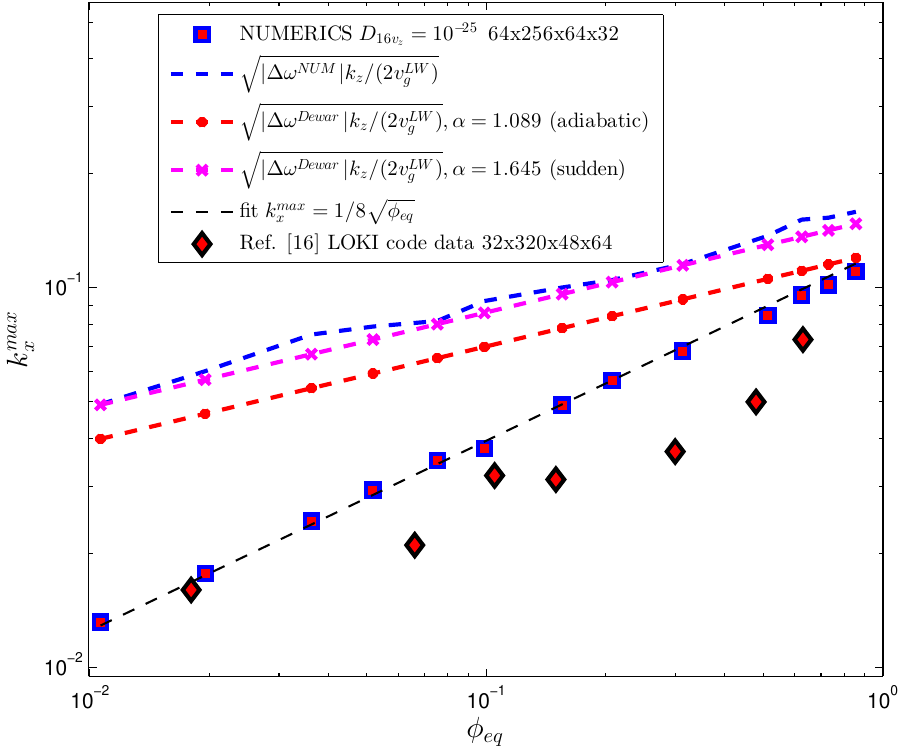}
\caption{(Color online) The wavenumber $k_x^{max}$ at which the growth rate reaches the maximum as a function of $\phi_{eq}$ for $k_z=0.425$, $v_{\varphi}=3.1$.}
\label{fig:kx_max_vs_phi_EPW_0425}
\end{figure}

The wavenumber $k_x^{max}$ at which the growth rate is maximum as a function of $\phi_{eq}$ is shown in Fig. \ref{fig:kx_max_vs_phi_EPW} together with the theoretical predictions given by Eq. \e{kmaxapprox} with different estimates for $\Delta\omega.$ For the group velocity $v_g$   in Eq. \e{kmaxapprox} we use the value $v_g=v_g^{LW}=1.26112...$ that is calculated using the liner LW dispersion relation for $k_z=0.35$. For $\Delta\omega$ in theoretical predictions we use Dewar's model as well as the measured $\Delta\omega^{NUM}$ for weak and strong pumps cases. None of the theoretical approximations predict $k_x^{max}$ well for small amplitudes $\phi_{eq}$. All of them predict $k_x^{max} \propto \phi_{eq}^{1/4}$ while from numerical results we see that $k_x^{max} \sim 0.1 \sqrt{\phi_{eq}}$. Absolute values of the measured $k_x^{max}$ differ from $\sqrt{|\Delta\omega^{NUM}|k_z/(2v_g^{LW})}$ of  Eq. \e{kmaxapprox} by a factor $\sim3$ at $\phi_{eq}=0.01$ and by factor $\sim2$ at $\phi_{eq}=0.1$. The data from Ref. \cite{BergerBrunnerBanksCohenWinjumPOP2015} exhibit a similar scaling, but absolute values of $k_x^{max}$ are smaller by $50\%$ in average. The measured $k_x^{max}$ for the strong pump are $10-20\%$ larger than for the weak pump and exhibit the same scaling in the whole range of amplitudes $\phi_{eq}$.

The same kind of simulations with weak pump are done for $k_z=0.425$ with  $\omega_{LW}(k_z=0.425)=1.3176$, $v_{\varphi}=3.1$ and  $v_g=v_g^{LW}(k_z=0.425)=1.304545...$. The results and a comparison with data from Ref. \cite{BergerBrunnerBanksCohenWinjumPOP2015,BanksPrivate2016} (when available) are given in Figs. \ref{fig:delta_W_vs_phi_EPW_0425}-\ref{fig:kx_max_vs_phi_EPW_0425}. In this case our measured frequency shift $|\omega^{NUM}|$ is close to $\alpha=1.645$ (sudden) case in Dewar's theory. In Fig. \ref{fig:G_max_vs_phi_EPW_0425} the approximation $\gamma^{max}\approx|\Delta\omega^{NUM}|/4$ works pretty well for $\phi_{eq}<0.5.$ The measured growth rates are within $20-25\%$ from the estimate and scale like $\gamma^{max} \varpropto \sqrt{\phi_{eq}}$. Also for amplitudes $\phi_{eq}>0.5$ the scaling changes and becomes $\gamma^{max} \varpropto \phi_{eq}$. The data form Ref. \cite{BergerBrunnerBanksCohenWinjumPOP2015,BanksPrivate2016} exhibit a similar behaviour regarding the scalings, but absolute values of $\gamma^{max}$ are approximately 2 times smaller. Using Dewar's approximation for $|\Delta\omega|$ we notice that our growth rates are close to $|\Delta\omega^{Dewar}|/4$ with $\alpha=1.645$ (sudden), whereas data growth rates from Ref. \cite{BergerBrunnerBanksCohenWinjumPOP2015,BanksPrivate2016} are close to the case of $\alpha=1.089$ (adiabatic) for small amplitudes. Unfortunately, the measured $\Delta\omega^{NUM}$ from Ref. \cite{BergerBrunnerBanksCohenWinjumPOP2015,BanksPrivate2016} were not available for comparison. For $k_x^{max}$ we clearly see that $k_x^{max} \sim 1/8 \sqrt{\phi_{eq}}$, so none of the theoretical approximations predict $k_x^{max}$ well.

\section{Comparison of transverse instability of nonlinear EPWs and BGK modes}
\label{sec:2DComparison}
Here we compare the transverse instability results for weakly pumped EPWs with $k_z=0.35$ found in Section \ref{sec:2DPumpedEPW}  with transverse instability of BGK modes from Part I of this series. Notice that all BGK-like modes (weakly pumped EPWs) with various amplitudes were obtained using the pumping frequency $\omega_{LW}(k_z=0.35)=1.22095\ldots$ and, respectively, $v_{\varphi}=3.488$, whereas the BGK modes for different amplitudes were constructed such that $v_{\varphi}=v_{\varphi}(\phi_{eq})$ according to the dispersion relation given by Eq. (22) of Part I.

\begin{figure}
\includegraphics[width=2.8in]{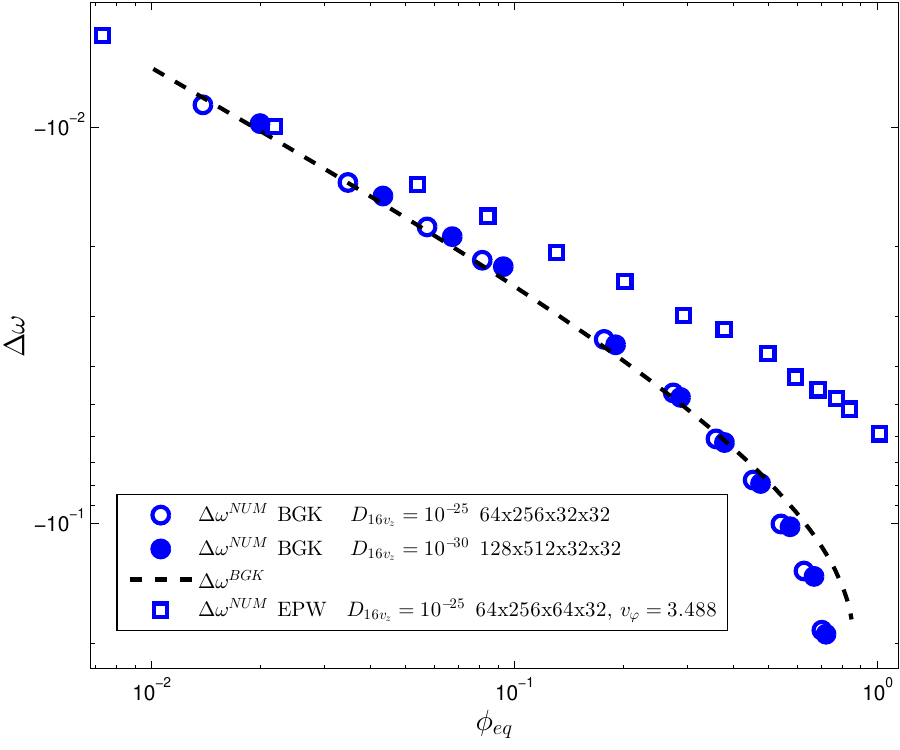}
\caption{(Color online) The nonlinear frequency shift $\Delta \omega$ as a function of $\phi_{eq}$ for both BGK modes and pumped EPWs (BGK-like modes) with $k_z=0.35$.}
\label{fig:delta_W_vs_phi_COMPARISON}
\end{figure}
\begin{figure}
\includegraphics[width=2.9in]{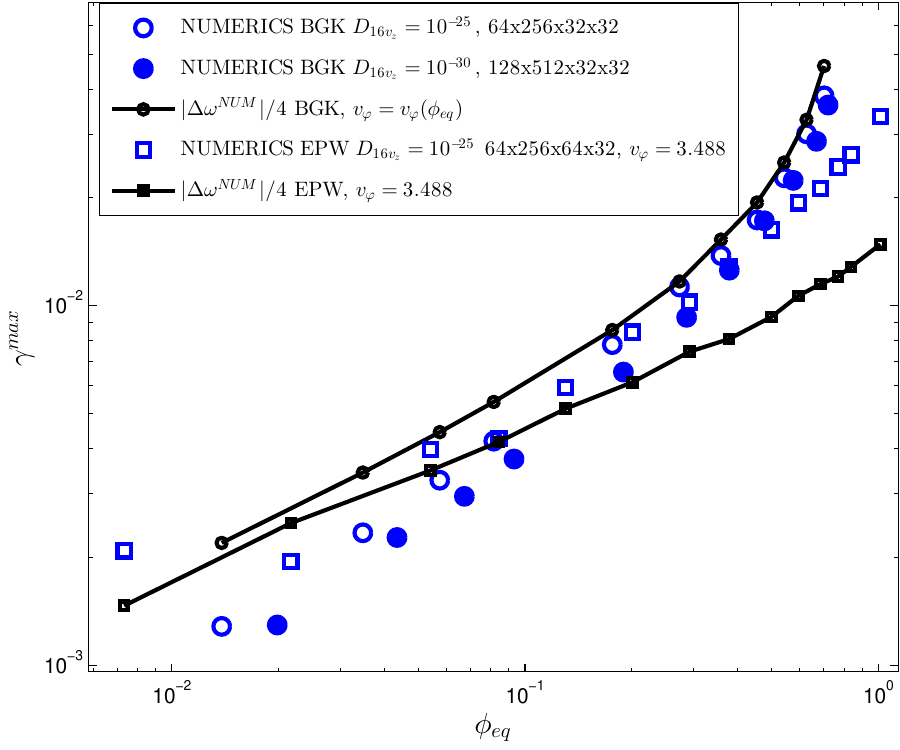}
\caption{(Color online) The maximum growth rate as a function of $\phi_{eq}$ for both BGK modes and pumped EPWs with $k_z=0.35$.}
\label{fig:G_max_vs_phi_COMPARISON}
\end{figure}
\begin{figure}
\includegraphics[width=2.9in]{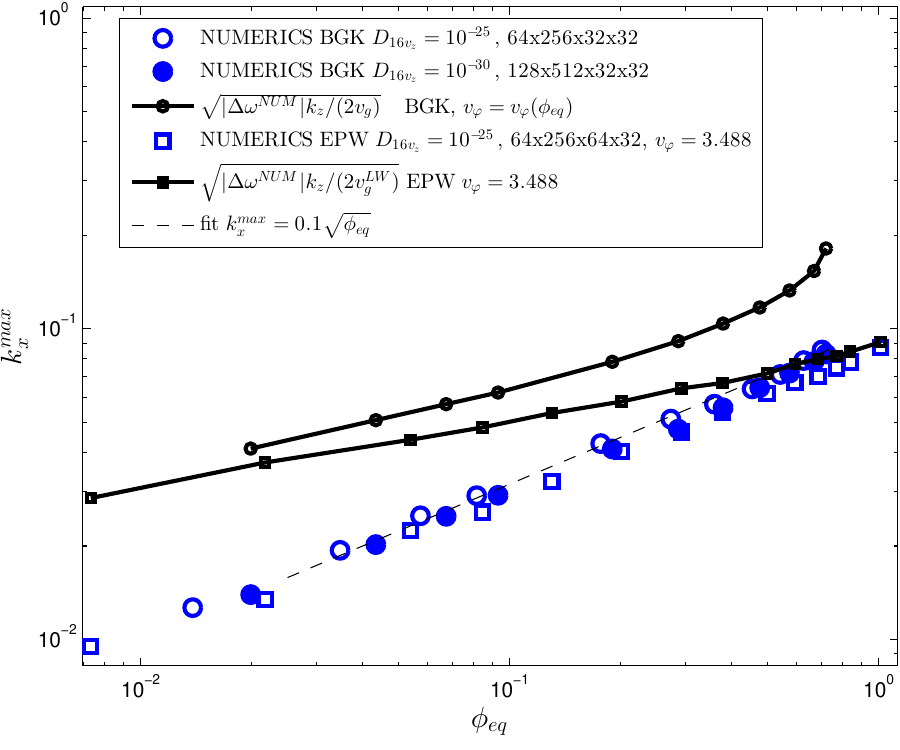}
\caption{(Color online) The wavenumber $k_x^{max}$ at which the growth rate reaches the maximum as a function of $\phi_{eq}$ for both BGK modes and pumped EPWs with $k_z=0.35$.}
\label{fig:kx_max_vs_phi_COMPARISON}
\end{figure}

Fig. \ref{fig:delta_W_vs_phi_COMPARISON} shows the nonlinear frequency shift obtained from both kinds of simulations. We can see that for small amplitudes $\phi_{eq}<0.05,$ the nonlinear frequency shift for both BGK and pumped EPWs basically coincides, whereas for higher amplitudes it changes its scaling for BGK modes and stays $\propto \sqrt{\phi_{eq}}$ for weakly pumped EPWs.

The maximum growth rate $\gamma^{max}$ (the maximum vs. $k_x$ for each fixed  $\phi_{eq}$) as a function of $\phi_{eq}$ is shown in Fig. \ref{fig:G_max_vs_phi_COMPARISON} together with the theoretical predictions given by $\gamma^{max}\approx|\Delta\omega^{NUM}|/4$ from Eq. \e{gammamaxsqrt}. We see that growth rates coincide for a wide range of amplitudes up to $\phi_{eq}<0.5$ despite the growing difference in the nonlinear frequency shift between these two kinds of waves in Fig. \ref{fig:delta_W_vs_phi_COMPARISON} (e.g. at $\phi_{eq}=0.5$, the BGK mode nonlinear frequency shift is twice larger than for  BGK-like pumped mode).

The wavenumber $k_x^{max}$ at which the growth rate is maximum as a function of $\phi_{eq}$ is shown in Fig. \ref{fig:kx_max_vs_phi_COMPARISON} together with the theoretical predictions given by $k_x^{max}\approx \sqrt{|\Delta\omega^{NUM}|k_z/(2v_g)}$ from Eq. \e{kmaxapprox}. We used BGK dispersion relation Eq. (22) of Part I to calculate $v_g$ for the comparison with BGK results and linear LW dispersion to calculate $v_g^{LW}$ for the comparison with pumped EPW results. We see that $k_x^{max}$ for these two classes of waves coincide for the whole range of amplitudes (up to $\phi_{eq}\thickapprox0.72$) and fit well to $k_x^{max} = 0.1 \sqrt{\phi_{eq}}$ law despite quite a big discrepancy with theoretical predictions.

These results suggest that the nonlinear frequency shift $\Delta\omega$ or the amplitude $\phi_{eq}$ are not sufficient to fully characterize the transverse instability of BGK and BGK-like modes. Perhaps the details of the phase space distribution function $f$ behaviour in the trapping region have to be taken into account which is however beyond the scope of this paper.

\begin{figure}
\includegraphics[width=2.9in]{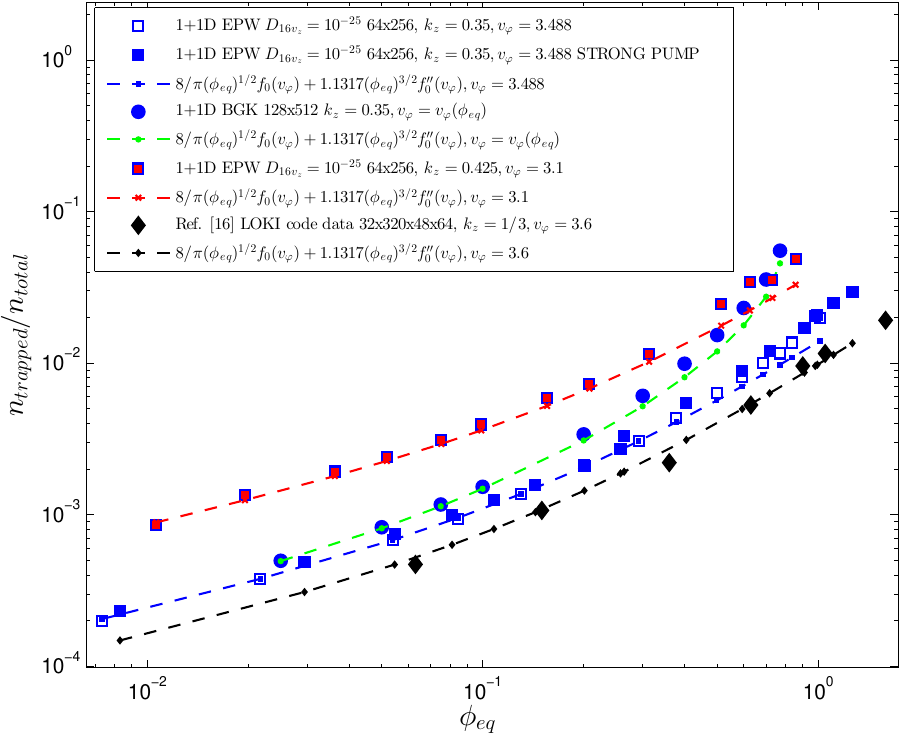}
\caption{(Color online) The fraction of trapped particles $n_{trapped}/n_{total}$ as a function of $\phi_{eq}$ for both BGK modes and pumped EPWs.}
\label{fig:n_trapped_vs_phi_eq_COMPARISON}
\end{figure}
\begin{figure}
\includegraphics[width=2.9in]{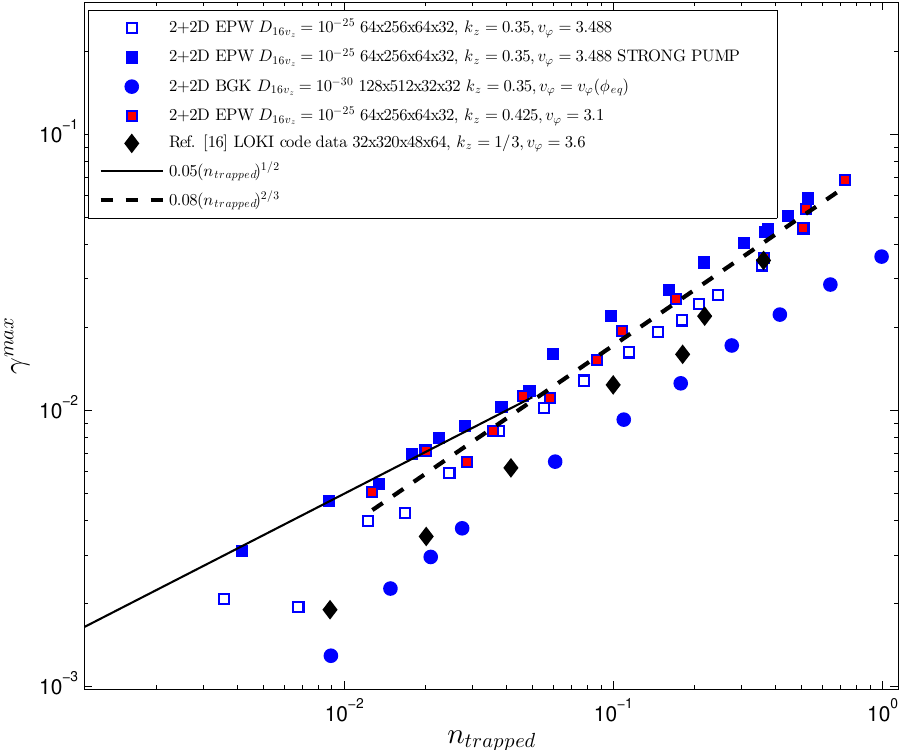}
\caption{(Color online) The maximum growth rate as a function of $n_{trapped}$ for both BGK modes and pumped EPWs.}
\label{fig:G_MAX_vs_n_trapped_COMPARISON}
\end{figure}

We also compared in Fig. \ref{fig:n_trapped_vs_phi_eq_COMPARISON} the fraction of trapped particles $n_{trapped}/n_{total}$ for all simulation data we obtained (marked with squares for the pumped EPWs with $k_z=0.35$ and $k_z=0.425$, circles for the BGK modes with $k_z=0.35$, and diamonds for the data from Refs.\cite{BergerBrunnerBanksCohenWinjumPOP2015},\cite{BanksPrivate2016} for EPWs with $k_z=1/3$) with the theoretical prediction (dashed lines with the corresponding markers) from Ref. \cite{RoseRussellPOP2001}:
\begin{eqnarray}  \label{eq:n_trapped}
\nonumber \frac{n_{trapped}}{n_{total}} &\approx& \frac{8}{\pi} (\phi_{eq})^{1/2} f_0(v_{\varphi}) + 1.1317(\phi_{eq})^{3/2} f_0''(v_{\varphi}), \\
n_{trapped}&=&\iint\limits_{W<\Phi_{max}} f(z,v_z)dv_zdz, \\
\nonumber n_{total}&=&\iint f(z,v_z)dv_zdz=L_z=\frac{2\pi}{k_z}.
\end{eqnarray}
It was  derived for  the BGK modes of Part I but we found   it to work really well for pumped EPWs also. Here $f_0$ is defined in Eq. \e{maxw1}. Eqs. \e{eq:n_trapped} take into account not only the leading order term ($\phi_{eq}^{1/2}$) approximation but the next order term ($\phi_{eq}^{3/2}$) as well. One can see in Fig. \ref{fig:n_trapped_vs_phi_eq_COMPARISON} that the data are within $10\%$ from the corresponding theoretical curves for all of our simulation with $\phi_{eq}\lesssim0.3$. Also EPWs with $k_z=0.35$ obtained using the strong pump exhibit $\approx10\%$ higher values of $n_{trapped}$ compared to EPWs obtained using weak pump. Notice that the EPW results with $k_z=0.35$ converge to the BGK results with $k_z=0.35$ in the limit $\phi_{eq}\rightarrow0$ as expected since the BGK waves were constructed as a finite-amplitude bifurcation of a linear LW.
For pumped EPWs, $n_{trapped}$ was calculated numerically from 1+1D simulations some time after the pump was switched off (typically $t=1000$). As EPW evolved in our simulations between $t=T_{off}$ and $t=1000,$ then $n_{trapped}$ would typically decreased only by $1-2\%$. For BGK modes, $n_{trapped}$ was calculated numerically after constructing 1+1D BGK solution analytically (no evolution). Also note that for the pumped EPW, $v_{\varphi}$ is the same (and given by LW dispersion relation) in all simulations with a particular $k_z$, while for BGK modes, $v_{\varphi}=v_{\varphi}(\phi_{eq})$ according to the dispersion relation given by Eq. (22) of Part I. We have not included into Fig. \ref{fig:n_trapped_vs_phi_eq_COMPARISON} the number of trapped particles for BGK modes obtained on a smaller resolution (64x256) as in Part I since the difference in $n_{trapped}$ values was less than $1\%$.

Also Fig. \ref{fig:G_MAX_vs_n_trapped_COMPARISON} shows the maximum growth rate $\gamma^{max}$ as a function of $n_{trapped}$ for both BGK modes and pumped EPWs. Even though it is hard to conclude anything regarding the scaling for the simulations with small amplitudes (left side of the graph) due to the larger numerical errors for $\gamma^{max}$ in these simulations, it appears that   the dependance $\gamma^{max}$ on $n_{trapped}$ has somewhat more universal scaling (somewhat close to $\gamma^{max} \propto (n_{trapped})^{2/3}$) for higher amplitudes compared to the dependance $\gamma^{max}$ on $\phi_{eq}$ in Figs. \ref{fig:G_max_vs_phi_EPW}, \ref{fig:G_max_vs_phi_EPW_0425} and \ref{fig:G_max_vs_phi_COMPARISON}, where the scaling changes from $\gamma^{max}\propto \sqrt{\phi_{eq}}$ to $\gamma^{max}\propto \phi_{eq}$. 

\section{CONCLUSION}
\label{sec:Conclusion}
We studied the filamentation of Langmuir wave in the kinetic regime $k\lambda_D\gtrsim0.2$ considering EPWs obtained by  pumping of the system by external electric potential. Weak and strong pumps are considered and compared. Performing direct 2+2D Vlasov-Poisson simulations of collisionless plasma we found that the maximal growth rates $\gamma^{max}$  for weakly pumped EPW are within $20-30\%$ from the theoretical prediction for small amplitudes ($\phi_{eq}<0.2$)  both for $k_z=0.35$ and $k_z=0.425$. Strongly pumped LWs  have higher filamentation grow rates. Also $\gamma^{max}$ for both types of pumping exhibits the proper scaling for small amplitudes of EPWs $\gamma^{max} \propto \sqrt{\phi_{eq}}$ while $k_x^{max} \propto \sqrt{\phi_{eq}}$ result remains to be explained theoretically since current theory (Eqs. \e{eq:dW_Dewar} and \e{kmaxapprox}) predicts $k_x^{max} \propto (\phi_{eq})^{1/4}$. Also it appears that the scaling $\gamma^{max} \propto (n_{trapped})^{2/3}$ might be somewhat more universal among pumped EPWs and BGK modes with various $k_z$ and amplitudes.

We found that both  BGK modes and weakly pumped BGK-like modes  have the same transverse instability growth rates for $k_z=0.35$  and peaked at the same wavenumber $k_x=k_x^{max}$ even though the electron phase space  distribution function $f(z,v_z,t)$ is not the same for these solutions as shown in Fig. \ref{fig:1D_Cross_sections}.  It suggests the universal mechanism for the kinetic saturation of stimulated Raman scatter in laser-plasma interaction experiments.

\begin{acknowledgments}
The authors thanks J. W. Banks and R. L. Berger for helpful discussion on 2D Vlasov simulations and for the data provided for the comparison with our simulations.
This work was supported by the National Science Foundation
under Grants No. PHY 1004118, PHY 1004110 and DMS-1412140.
Simulations were performed at the Center for Advanced Research Computing (CARC) at the University of New Mexico and the Texas Advanced Computing Center (TACC) which was supported by National Science Foundation Grant ACI-1053575.
\end{acknowledgments}



%

\end{document}